\documentclass[journal=jpca,manuscript=article]{achemso}

\usepackage[version=4]{mhchem} 
\usepackage{tikz}
\usepackage{circuitikz}
\usepackage{booktabs}
\usepackage{longtable}
\usepackage{siunitx}
\usepackage{multirow}
\usepackage{threeparttablex}
\usepackage{caption}
\usepackage{subcaption}
\usepackage{rotating}
\usepackage{blindtext}
\usepackage{chemmacros}
\DeclareSIUnit{\hartree}{E_h}

\author{Jacopo Lupi}
\affiliation[TCD]
{School of Physics, Trinity College Dublin, Dublin 2, Ireland}
\alsoaffiliation[AMBER]{AMBER, Advance Materials and BioEngineering Research Centre}
\email{jacopo.lupi@tcd.ie}
\author{Leandro Ayarde-Henríquez}
\affiliation[TCD]
{School of Physics, Trinity College Dublin, Dublin 2, Ireland}
\alsoaffiliation[AMBER]{AMBER, Advance Materials and BioEngineering Research Centre}
\author{Mark Kelly}
\affiliation[TCD]
{School of Physics, Trinity College Dublin, Dublin 2, Ireland}
\alsoaffiliation[AMBER]{AMBER, Advance Materials and BioEngineering Research Centre}
\author{Stephen Dooley}
\email{stephen.dooley@tcd.ie}
\affiliation[TCD]
{School of Physics, Trinity College Dublin, Dublin 2, Ireland}
\alsoaffiliation[AMBER]{AMBER, Advance Materials and BioEngineering Research Centre}

\title[Xylopyranose modelling]
  {Ab Initio and Kinetic Modelling of $\beta$-D-xylopyranose Under Fast Pyrolysis Conditions} 

\abbreviations{DFT,QM,KM,PES}

\begin{document}

\begin{tocentry}
\includegraphics[width=1.05\linewidth]{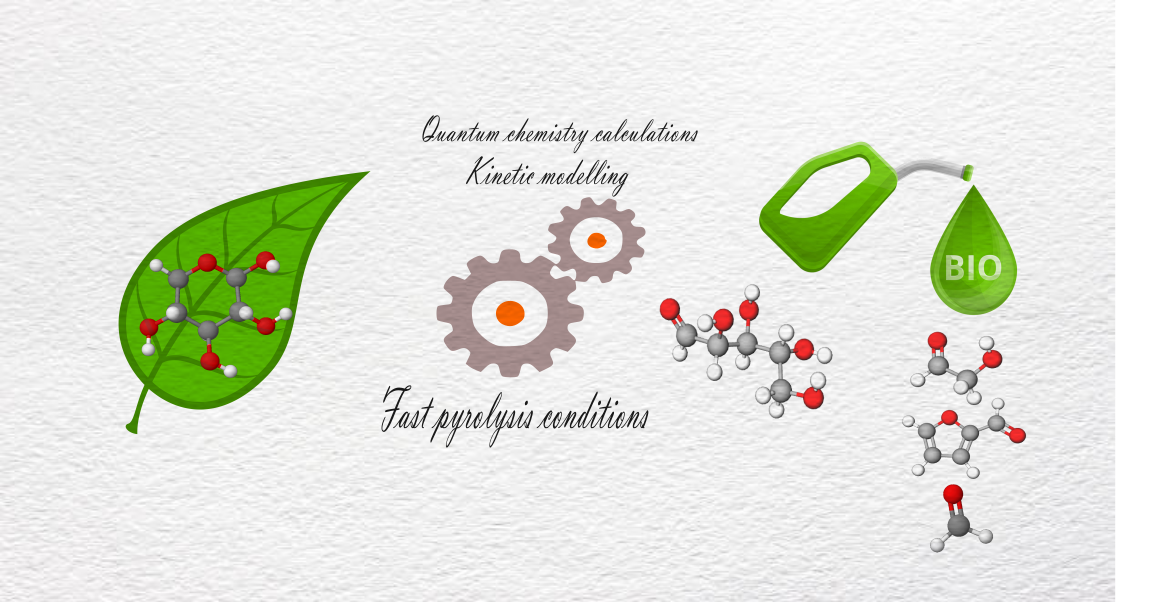}
\end{tocentry}

\begin{abstract}
 Lignocellulosic biomass is an abundant renewable resource that can be upgraded to chemical and fuel products through a range of thermal conversion processes. Fast pyrolysis is a promising technology that uses high temperatures and fast heating rates to convert lignocellulose into bio-oils in high yields in the absence of oxygen. Hemicellulose is one of three major components of lignocellulosic biomass and is a highly branched heteropolymer structure made of pentose, hexose sugars, and sugar acids.
In this study, $\beta$-D-xylopyranose is proposed as a model structural motif for the essential chemical structure of hemicellulose. The gas-phase pyrolytic reactivity of $\beta$-D-xylopyranose is thoroughly investigated using computational strategies rooted in quantum chemistry. In particular, its thermal degradation potential energy surfaces are computed employing Minnesota global hybrid functional M06-2X in conjunction with 6-311++G(d,p) Pople basis set. Electronic energies are further refined by performing DLPNO-CCSD(T)-F12 single point calculations on top of M06-2X geometries using cc-pVTZ-F12 basis set. Conformational analysis for minima and transition states is performed with state-of-art semiempirical quantum chemical methods coupled with metadynamics simulations. Key thermodynamic quantities (free energies, barrier heights, enthalpies of formation, and heat capacities) are computed. Rate coefficients for the initial steps of thermal decomposition are computed by means of reaction rate theory. For the first time, a detailed elementary reaction kinetic model for $\beta$-D-xylopyranose is developed by utilizing the thermodynamic and kinetic information acquired from the aforementioned calculations. This model specifically targets the initial stages of $\beta$-D-xylopyranose pyrolysis, aiming to gain a deeper understanding of its reaction kinetics. This approach establishes a systematic strategy for exploring reactive pathways, evaluating competing parallel reactions, and selectively accepting or discarding pathways based on the analysis. The findings suggest that acyclic D-xylose plays a significant role as an intermediary in the production of key pyrolytic compounds during the pyrolysis of xylose. These compounds include furfural, anhydro-D-xylopyranose, glycolaldehyde, and dihydrofuran-3(2H)-one.
\end{abstract}

\section{Introduction}
Biomass pyrolysis refers to the thermal decomposition of organic matter, such as agricultural waste and wood, in the absence of oxygen to produce biofuels, such as bio-oil, charcoal, and biogas. As the demand for energy grows, biomass pyrolysis has emerged as a promising source of clean and sustainable energy. This is because it produces valuable biofuels and has the potential to reduce greenhouse gas emissions by replacing fossil fuels while also minimizing waste. For example, the European Union has set targets to increase the use of renewable energy and reduce greenhouse gas emissions \cite{EUGreenDeal}. This has led to increased investment in biomass pyrolysis technologies, as well as the development of policies and incentives to support their deployment.

In particular, non-food, non-feed lignocellulosic plant matter is a highly desirable form of biomass with numerous advantages. Its abundance, widespread availability, and cost-effectiveness make it an attractive option for various applications, owing to its prevalence in crop residues, woody plants, and grasses. Chemicals, fuels, and energy must be produced from lignocellulosic plant matter in order to decarbonize our global economies, and, to be successful, these lignocellulose-derived products must compete in quality and price with fossil-derived products. Lignocellulose is formed by three main components, namely, cellulose, hemicellulose, and lignin.

Hemicellulose is the second most abundant component of lignocellulosic biomass and is a group of cell wall polysaccharides distinct from cellulose and pectin. Some examples of hemicelluloses include xylans (such as $\beta$-1,4-D-xylan, $\beta$-1,3-D-xylan, $\beta$-1,3;1,4-D-xylan, 4-O-methyl-D-glucorono-D-xylan, glucuronoxylan, arabinoxylan, glucuronoarabinoxylan, and arabinoglucuronoxylan), mannans (like homomannan, galactomannan, glucomannan, and galactoglucomannan), xyloglucans, $\beta$-1,3;1,4-glucans, and galactans (such as sulfated galactans, arabinogalactans). The main hemicellulose polysaccharides found in hardwood, softwood, and herbaceous biomass are 4-O-methylglucoronoxylans, galactoglucomannan, and arabinoxylans, respectively. A comprehensive review on hemicellulose pyrolysis has been published in 2017 by Broadbelt and collaborators.\cite{zhou2017}

Zhou \emph{et al.}\cite{zhou2018} proposed one of the first mechanistic models for fast hemicellulose pyrolysis. This model, based on the experimentally characterized composition and linkages of extracted hemicellulose, rationalizes more than 500 reactions involved in hemicellulose decomposition and the formation of various pyrolysis products using the reaction family approach.  Additionally, the mechanistic model was extended to simulate fast pyrolysis of native hemicellulose from corn stalks. Their results show that native hemicellulose yields more char, gaseous species, acetol, and acetic acid compared to corn stalks extracted hemicellulose, while also yielding lower amounts of certain products. The presence and distribution of acetyl groups in hemicellulose significantly influence the pyrolysis product distribution, suggesting the potential impact of biomass pretreatment on pyrolysis outcomes.

Integrating physicochemical information robustly and coherently is crucial to enhancing and expanding kinetic models for lignocellulose pyrolysis. This valuable data can be obtained through computational chemistry techniques. Particularly, quantum chemistry modelling can give many details at the atomic/molecular level. It is usually performed to simulate the pyrolysis of the individual principal biomass components, interactions within/between these materials, and catalytic pyrolysis with various catalysts. On the basis of the theoretical models in quantum chemistry modelling, geometric structures, transition states, intermediates, corresponding electron transfers, orbital interactions, energetics, and other important information involved in the pyrolysis process can be obtained. Quantum chemistry modelling is playing increasingly important roles in the lignocellulosic biomass pyrolysis mechanism studies due to advancement of theoretical methods and computer hardware as well as the experimental limitations. In the literature, many reviews on the experimental studies of lignocellulosic biomass pyrolysis have been reported, and the topics cover the development of pyrolysis techniques,\cite{Mettler2012,Pires2019} reactors,\cite{Nunez2017} reaction conditions,\cite{Kan2016} selective preparation of high-grade bio-oil and value-added chemicals.\cite{Kabir2017,Baruah2018} As for the pyrolysis mechanism of biomass, most of the reviews focus on the pyrolysis kinetics, especially on global kinetic models.\cite{Burnham2015,Murillo2017,Hameed2019}.

However, most literature on biomass pyrolysis using quantum chemical modelling has been found to have a low-standard quantum chemical bias due to the absence of well-established and robust protocols in the field. The real danger lies in the scientific community of “non-theoretical chemists” persisting in using outdated functionals (such as B3LYP) simply due to their established history rather than their demonstrated effectiveness. This issue has been highlighted by Grimme and his colleagues in their recent works \cite{Bursch2022,Goerigk2017}. This low-quality quantum chemical modelling can introduce a significant level of uncertainty in computed properties (in particular geometries and reaction barriers), especially when these data are incorporated into global kinetic schemes to reproduce pyrolysis experiments. Therefore, developing new, more physically based protocols for accurate and cost-effective characterization of lignocellulosic pyrolytic systems is essential.

Numerous studies have been conducted using quantum chemistry to investigate the pyrolytic pathways of xylopyranose. However, these studies lack the desired level of comprehensiveness and computational accuracy. For instance, Wang \emph{et al.}\cite{wang2015theoretical} explored furfural formation, an important biofuel precursor, in both gas and water environments using B3LYP/6-31++G(d,p) and CPCM solvation models. They identified xylulose as an intermediate leading to furfural formation and proposed a feasible pathway involving hydrogen migrations and dehydrations, with solvent effects playing a significant role in stabilizing reactants and transition states.

In another work, Huang \emph{et al.} \cite{Huang2016} used M06-2X/6-31++G(d,p) to examine xylopyranose pyrolysis pathways, highlighting reactions leading to acyclic containing-carbonyl isomer (xylose) through immediate ring-opening, as well as pathways involving dehydration reactions. The major products obtained were low molecular weight compounds, including glycolaldehyde, 2-furaldehyde, acetaldehyde, and acetone, while other competitive products like formaldehyde, formic acid, acetic acid, \ce{CO2}, \ce{CH4}, acetol, and pyranone were also identified.

In 2019, Hu \emph{et al.}\cite{Hu2019} conducted a comprehensive investigation of xylopyranose thermal decomposition, employing analytical pyrolysis-gas chromatography/mass spectrometry (Py-GC/MS) and quantum chemistry calculations using B3LYP-D3/6-311G(d,p). They studied xylopyranose, its dimer, and xylan, finding that acyclic xylose played a key role as an intermediate in the formation of major pyrolytic products, such as 1,4-anhydro-D-xylopyranose, furfural, and glycolaldehyde during xylose pyrolysis.

Despite the considerable efforts made by multiple authors to study the pyrolytic pathways of $\beta$-D-xylopyranose using quantum chemistry, their investigations were limited by the use of an unsatisfactory quantum chemical level of theory and the absence of a detailed kinetic model to explore the interplay between thermodynamics and kinetics. As a result, in this study, the intention is to revisit the $\beta$-D-xylopyranose pyrolytic reactions, building upon the existing literature but employing a higher level of theory that is both robust and well-established.

The computational strategy employed comprises four fundamental components: (i) the characterization of stationary points within the reactive potential energy surface using the density functional theory; (ii) the refinement of electronic energies, vibrational zero-point energies, and determination of thermodynamic quantities; (iii) the determination of rate constants through the application of reaction rate theory; and (iv) the construction of a kinetic model that integrates thermodynamic and kinetic aspects to analyze their mutual influence. A perspective on a general computational strategy to iteratively investigate biomass-related chemical systems is also sketched.
The manuscript is organized as follows: firstly, a comprehensive exposition of the computational methods utilized in this paper is provided. Subsequent sections include the presentation of the computed reaction mechanism and an accompanying analysis of the associated thermodynamic properties. Additionally, the discussion encompasses the rate constants calculated via Transition State Theory. Finally, the outcomes of the kinetic model simulations are elucidated. The last section is reserved for summarizing the findings and outlining potential future directions.

\section{Computational methodology}
\label{compmeth}
\subsection{Electronic structure calculations}
The potential energy surfaces of $\beta$-D-xylopyranose thermal degradation reaction mechanism have been investigated through high-level quantum chemical calculations.

On the grounds of its well-known robustness \cite{Goerigk2017,Bursch2022,Hu2020}, geometry optimizations and harmonic frequencies of reactants, transition states, intermediates, and products along the reaction pathways were obtained by the M06-2X \cite{Zhao2008} global hybrid density functional in conjunction with the 6-311++G(d,p) basis set.\cite{Krishnan1980,Clark1983} The stationary points on the reaction pathways were characterized as minima (reactants, intermediates, and products) and saddle points (transition states) based on vibrational frequency calculations. The transition states obtained were further confirmed using intrinsic reaction coordinate (IRC) scans at the same levels of theory. 

Single point energy calculations were conducted on top of DFT geometries to further refine electronic energies at DLPNO-CCSD(T)\cite{dlpnocc} (with F12 explicit correlation correction)\cite{dlpnoccf12} level of theory in conjunction with cc-pVTZ-F12 basis set.\cite{peterson08} DLPNO calculations were performed employing the ORCA code\cite{orca5} and using tightPNO cut-off.

The neglect of anharmonicity tends to overestimate the vibrational zero-point energy (ZPE).\cite{abinitiopople} Therefore, in order to obtain accurate zero-point energies, avoiding the calculation of perturbative anharmonic corrections, ZPE and frequencies are scaled by 0.970.\cite{alecu2010,truhlarscaling} Quasi-Harmonic entropies were calculated employing Grimme's approximation. Thermochemical analysis was conducted using two codes: GoodVibes\cite{goodvibes} and Shermo.\cite{LU2021113249} The former code was used for enthalpy, entropy and free energy calculations, while the Shermo code was specifically utilized for calculating constant pressure heat capacities.
Given the high flexibility of the molecules under investigation, conformational analysis for minima and transition states is performed with the CREST program \cite{Pracht2020} in order to provide starting guesses for the potential energy surfaces calculations. The code couples state-of-art semiempirical quantum chemical methods (GFNn-xTB) \cite{Bannwarth2021} with metadynamics simulations \cite{Grimme2019}.
All DFT geometry optimization and frequency calculations have been performed using the Gaussian code \cite{g16}.

\subsection{Enthalpies of formation calculation}
Taking xylopyranose as an example, the enthalpies of formation ($\Delta _fH$) of the molecular species were calculated using the enthalpies of atomization as\cite{atomization}
\begin{equation*}
\begin{split}
    \Delta _f&H^{\standardstate}(\ce{C5H10O5}) =\Delta^{theo} _rH^{at}+\sum_{\textrm{atoms}}\Delta _{f}^{exp}H^{\standardstate} \\
    & =[H(\ce{C5H10O5})-5H(\ce{O})-5H(\ce{C})-10H(\ce{H})]^{theo} \\
    & + 5\Delta _{f}^{exp}H^{\standardstate}(\textrm{O})+5\Delta _{f}^{exp}H^{\standardstate}(\textrm{C})+10\Delta _{f}^{exp}H^{\standardstate}(\textrm{H}).
    \end{split}
\end{equation*}

In this expression, \textit{theo} refers to enthalpies calculated theoretically and \textit{exp} to ``experimental" values. The subscripts \textit{f} and \textit{r} refer to formation and reaction respectively, calculated or measured at 298.15 K. 
Atoms are considered in their standard state and their enthalpies of formation are taken from the Active Thermochemical Tables (ATcT)\cite{ruscic2004,brankoRuscic_2005,ruscic2022active}, namely H($^2$S)=\SI{217.998}{\kilo\joule\per\mol}; C($^3$P)=\SI{716.881}{\kilo\joule\per\mol}; O($^3$P)=\SI{249.229}{\kilo\joule\per\mol}.

A final note is deserved. The limitations in accuracy of this method are known in comparison to error-cancelling methods like isodesmic reaction ones (i.e., those in which the number and type of different bonds in the species in the right-hand side and left-hand side of the chemical equation match as closely as possible).\cite{ghosh2019,ventura2022} Nevertheless, the atomization method has been chosen due to its simplicity and immediate applicability to the large number of molecules requiring analysis.

\subsection{Rate constants calculation}
High-pressure limit rate constants are determined by the conventional transition state theory (TST) within the rigid-rotor harmonic-oscillator (RRHO) approximation, be expressed by;
\begin{equation}
    k(T)=\kappa(T)\frac{m^\ddagger\sigma_{ext}}{m\sigma^\ddagger _{ext}}\frac{k_BT}{h}\frac{Z^\ddagger}{Z}\exp\left(-\frac{E_0}{k_BT}\right).
\end{equation}
Here $\kappa(T)$ is the transmission coefficient, which accounts for tunneling as well as nonclassical reflection effects using the one-dimensional asymmetric Eckart model. $m^\ddagger$ and $m$ denote the number of enantiomers for the transition state and reactants, respectively, while $\sigma_{ext}$ and $\sigma$ refer to the symmetry numbers for external rotation of these entities. The partition functions for the transition state and reactants are denoted by $Z^\ddagger$ and $Z$, respectively. Furthermore, the constants $h$, $T$, and $k_B$ represent Planck's constant, temperature, and Boltzmann's constant, respectively. The term $E_0$ represents the barrier height, including the zero-point energy. 
To model their temperature dependence, the rate constants at different temperatures have
been fitted to Arrhenius equation:
\begin{equation}\label{eq:arrhenius}
    k(T)=A\exp{\left(-\frac{E_a}{RT}\right)}.
\end{equation}
The TST calculations have been performed using MESS code by Georgievskii \emph{et al.} \cite{Georgievskii2013}. 
\subsection{Kinetic modelling}
A kinetic model is built upon the chemical-physical informations obtained from thermodynamics and kinetics calculations. This includes providing Arrhenius pre-exponent ($A$) and activation energy ($E_a$) for each reaction, as well as information on the molar heat capacity at constant pressure, molar enthalpy of formation, and molar entropy of each chemical species. 

Specifically, in this work, two different kinetic models are built to assess the competition between initial competitive pathways. The first kinetic model is comprised of the following reactions:
\begin{align*}
& \ce{Xylopyranose <-> Xylose}, \\
& \ce{Xylopyranose <-> TH-C-OH{2}-c + H2O}, \\
& \ce{Xylopyranose <-> TH-C-OH{2}-t + H2O}, \\
& \ce{Xylopyranose <-> AXP1 + H2O}, \\
& \ce{Xylopyranose <-> AXP_{2-1} + H2O}, \\
& \ce{Xylopyranose <-> AXP_{2-3} + H2O}, \\
& \ce{Xylopyranose <-> AXP_{3-2} + H2O}, \\
& \ce{Xylopyranose <-> AXP_{3-4} + H2O}, \\
& \ce{Xylopyranose <-> AXP_{4-3} + H2O}, \\
& \ce{Xylopyranose <-> AXP_{4-5} + H2O}.
\end{align*}
These reactions specifically describe the initial stage of thermal decomposition observed in xylopyranose. The second kinetic model is comprised of the following reactions:
\begin{align*}
& \ce{Xylose <-> A{1} + H2O}, \\
& \ce{Xylose <-> B{1}}, \\
& \ce{Xylose <-> C{1} + C{2}}, \\
& \ce{Xylose <-> D{1}}, \\
\end{align*}
which are the series of reactions that xylose can undergo in the subsequent thermal decomposition steps. The significance of this naming convention and the justification for choosing these particular reaction sets will be clarified in the forthcoming discussion within the Results section.

For convenience, the thermodynamic properties for each chemical species are expressed in the NASA polynomial format as follows:
\begin{equation}
    \frac{\hat{C_p}^\standardstate(T)}{R}=a_0+a_1T+a_2T^2+a_3T^3+a_4T^4,
\end{equation}
\begin{equation}
    \frac{\hat{H}_f^\standardstate(T)}{RT}=a_0+\frac{a_1}{2}T+\frac{a_2}{3}T^2+\frac{a_3}{4}T^3+\frac{a_4}{5}T^4+\frac{a_5}{T},
\end{equation}
\begin{equation}
    \frac{\hat{S}^\standardstate(T)}{R}=a_0\ln T+a_1T+\frac{a_2}{2}T^2+\frac{a_3}{3}T^3+\frac{a_4}{4}T^4+a_6.
\end{equation}
This format allows for the expression of the temperature dependence of the thermodynamic properties of each chemical species. These $a_x$ coefficients must be provided in the kinetic model for each species in the system. Hence, in this work, the molar heat capacity at constant pressure, molar enthalpy of formation, and molar entropy of each chemical species are computed at a range of temperatures and fitted to the NASA format using the THERM code.\cite{therm} All reactions are reversible and hence the equilibrium rate constants are computed using the thermodynamic properties in conjunction with the unidirectional rate constants, computed as described above.

\subsection{Reactor model}
To model the thermal decomposition of xylose, the present study employs Cantera,\cite{cantera} an open-source chemical kinetic numerical solver. To perform a kinetic simulation using Cantera, a kinetic model must be provided, and the reactor physics must be declared.
Zero-dimensional (0-D) simulations are performed to model the time evolution of xylopyranose in a constant-pressure reactor at a set of operating conditions. Two types of simulation are performed. In the first one, the starting temperature is set at \SI{673.15}{\kelvin}, \SI{773.15}{\kelvin} and \SI{1073.15}{\kelvin} and the kinetic model evolves the species mole fractions as a function of time. 

In the second, the evolution of the mixture as the temperature of the reactor increases at a fixed heating rate of \SI{30}{\kelvin\per\minute} is modelled. This simulation is a representation of a thermogravimetric analysis (TGA) experiment which is a powerful technique for examination of the thermal degradation behaviour of a sample. Initial conditions of the reactor are set to one atmosphere and 100\% mass fraction of xylopyranose. All the species are in the gas-phase.

\section{Results and discussion}
\subsection{Molecular thermodynamics}
\subsubsection{Thermodynamics of initial decomposition pathways of xylopyranose}
\begin{figure}[htbp]
\centering
\includegraphics[width=0.7\columnwidth]{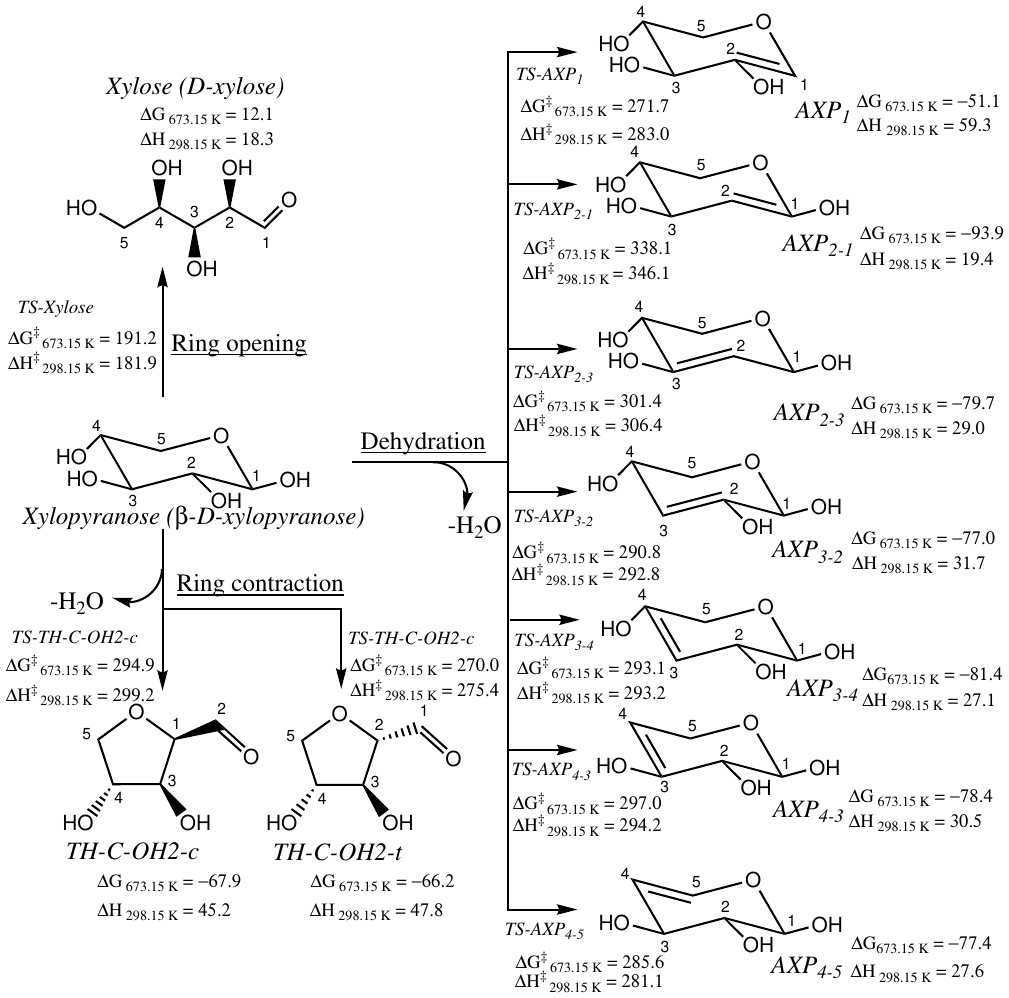} 
\caption{Initial steps of xylopyranose thermal degradation: ring-opening, ring-contraction and dehydration. Enthalpies and free energies are in \SI{}{\kilo\joule\per\mol}.} 
\label{fig:initialsteps}
\end{figure}

$\beta$-D-xylopyranose (\textbf{xylopyranose} hereafter) can undergo three reaction pathways, as already recognized in literature\cite{Huang2016,Hu2019}, namely ring opening, ring contraction and water elimination (dehydration) reactions. Figure \ref{fig:initialsteps} illustrates these reactions, with the energies summarized in Table \ref{tab:relativethermo}.

The ring opening reaction of \textbf{xylopyranose} leads to the formation of acyclic D-xylose (\textbf{xylose} hereafter).

The ring-contraction reactions occur in two ways through \textbf{TS-TH-C-OH2-c} and \textbf{TS-TH-C-OH2-t}, resulting in the formation of five-membered intermediates, \textbf{TH-C-OH2-c} and \textbf{TH-C-OH2-t}, i.e. dihydroxytetrahydrofuran-2-carbaldehyde. In the first of the two, the C2-C3 bond breaks while simultaneously forming the C1-C3 bond. In the second one, the C2-O bond forms as the C1-O bond cleavages. Regarding the dehydration reactions, xylopyranose can dehydrate at four sites (1 to 4), leading to seven possible anhydroxylopyranose (AXP) products: \textbf{\ce{AXP1}}, \textbf{\ce{AXP_{2-1}}}, \textbf{\ce{AXP_{2-3}}}, \textbf{\ce{AXP_{3-2}}}, \textbf{\ce{AXP_{3-4}}}, \textbf{\ce{AXP_{4-3}}} and \textbf{\ce{AXP_{4-5}}}. 
Examination of Table \ref{tab:relativethermo} reveals that the ring opening reaction exhibits the lowest barrier height, namely \SI{183.8}{\kilo\joule\per\mol}. Finite temperature effects are observed to slightly decrease the enthalpic barrier by approximately \SI{2}{\kilo\joule\per\mol} at \SI{298.15}{\kelvin}. However, due to entropic effects, the barrier is increased to \SI{191.2}{\kilo\joule\per\mol} at \SI{673.15}{\kelvin}. Notably, among the initial decomposition steps, the formation of \textbf{xylose} product stands out as the sole endergonic process, with a calculated free energy of reaction of \SI{12.1}{\kilo\joule\per\mol}.
\begin{table}[ht]
\centering
\caption{Relative enthalpies and free energies at \SI{0}{\kelvin}, \SI{298.15}{\kelvin}, and \SI{673.15}{\kelvin} at DLPNO-CCSD(T)-F12/cc-pVTZ-F12//M06-2X/6-311++G(d,p) level of theory for the species involved in xylopyranose initial thermal decomposition steps. Pressure is set to 1 atm. Energies in \SI{}{\kilo\joule\per\mol}.}
\label{tab:relativethermo}
\begin{tabular}{@{}lcccc@{}}
\toprule
          & $\Delta H_{\SI{0}{\kelvin}}$ & $\Delta H_{\SI{298.15}{\kelvin}}$ & $\Delta G_{\SI{298.15}{\kelvin}}$ & $\Delta G_{\SI{673.15}{\kelvin}}$ \\ \midrule
Xylopyranose                  & 0.0      & 0.0           & 0.0           & 0.0           \\
\ce{TS{-}Xylose}        & 183.8    & 181.9         & 185.5         & 191.2         \\
Xylose                   & 18.2     & 18.3          & 15.5          & 12.1          \\
\ce{TS{-}TH-C-OH{2}-c}       & 299.3    & 299.2         & 297.1         & 294.9         \\
\ce{TS{-}TH-C-OH{2}-t}       & 274.7    & 275.4         & 273.1         & 270.0         \\
\ce{TH-C-OH{2}-c + H2O}      & 39.2     & 45.2          & -5.0          & -67.9         \\
\ce{TH-C-OH{2}-t + H2O}      & 41.7     & 47.8          & -2.8          & -66.2         \\
\ce{TS{-}AXP1}        & 281.9    & 283.0         & 278.2         & 271.7         \\
\ce{TS{-}AXP_{2-1}}        & 345.6    & 346.1         & 342.8         & 338.1         \\
\ce{TS{-}AXP_{2-3}}        & 306.4    & 306.4         & 304.3         & 301.4         \\
\ce{TS{-}AXP_{3-2}}        & 293.0    & 292.8         & 292.1         & 290.8         \\
\ce{TS{-}AXP_{3-4}}        & 292.9    & 293.2         & 293.5         & 293.1         \\
\ce{TS{-}AXP_{4-3}}        & 294.5    & 294.2         & 295.8         & 297.0         \\
\ce{TS{-}AXP_{4-5}}        & 281.6    & 281.1         & 283.3         & 285.6         \\
\ce{AXP1 + H2O}       & 52.7     & 59.3          & 10.8          & -51.1         \\ 
\ce{AXP_{2-1} + H2O}  & 12.6     & 19.4          & -30.4         & -93.9         \\
\ce{AXP_{2-3} + H2O}  & 23.1     & 29.0          & -18.9         & -79.7         \\
\ce{AXP_{3-2} + H2O}  & 25.8     & 31.7          & -16.1         & -77.0         \\
\ce{AXP_{3-4} + H2O}  & 21.1     & 27.1          & -20.6         & -81.4         \\
\ce{AXP_{4-3} + H2O}  & 24.2     & 30.5          & -17.4         & -78.4         \\
\ce{AXP_{4-5} + H2O}  & 21.7     & 27.6          & -18.6         & -77.4         \\ \bottomrule
\end{tabular}%

\end{table}
Table \ref{tab:xylopyranosebarriers} presents an in-depth analysis of the forward and reverse barriers associated with the initial processes. A striking observation is that the enthalpic reverse barriers consistently exhibit lower values compared to their direct counterparts, both at \SI{0}{\kelvin} and \SI{298.15}{\kelvin}. However, this trend is reversed at \SI{673.15}{\kelvin}, considering the influence of entropic effects on free energies. From an entropic perspective, bimolecular reactions are less favored, resulting in increased free energy barriers. It is worth noting that the ring opening reaction stands as the only exception, as it follows an unimolecular pathway in both directions. 

\begin{table}[htbp]
\centering
\caption{Barrier heights at \SI{0}{\kelvin}, \SI{298.15}{\kelvin}, and \SI{673.15}{\kelvin} for the single-step reactions of xylopyranose initial thermal decomposition. Computed at DLPNO-CCSD(T)-F12/cc-pVTZ-F12//M06-2X/6-311++G(d,p) level of theory. Energies in \SI{}{\kilo\joule\per\mol}.}
\label{tab:xylopyranosebarriers}
\resizebox{\textwidth}{!}{%
\begin{tabular}{@{}lcccc@{}}
\toprule
                       & $\Delta H^\ddagger_{\SI{0}{\kelvin}}$ & $\Delta H^\ddagger_{\SI{298.15}{\kelvin}}$ & $\Delta G^\ddagger_{\SI{298.15}{\kelvin}}$ & $\Delta G^\ddagger_{\SI{673.15}{\kelvin}}$ \\ \midrule
                                        &\multicolumn{4}{c}{forward/reverse}            \\\midrule
\ce{Xylopyranose <-> Xylose}              &183.8/165.7          &181.9/163.7               &185.5/170.0               &191.2/179.1                  \\
\ce{Xylopyranose <-> TH-C-OH{2}-c + H2O}      &299.3/260.1          &299.2/254.0               &297.1/302.1               &294.9/362.8                  \\
\ce{Xylopyranose <-> TH-C-OH{2}-t + H2O}      &274.7/233.0          &275.4/227.6               &273.1/275.9               &270.0/336.2                  \\
\ce{Xylopyranose <-> AXP1 + H2O}       &281.9/229.2          &283.0/223.7               &278.2/267.4               &271.7/322.7                  \\ 
\ce{Xylopyranose <-> AXP_{2-1} + H2O}  &345.6/333.0          &346.1/326.8               &342.8/373.3               &338.1/432.0                  \\
\ce{Xylopyranose <-> AXP_{2-3} + H2O}  &306.4/283.3          &306.4/277.4               &304.3/323.2               &301.4/381.0                  \\
\ce{Xylopyranose <-> AXP_{3-2} + H2O}  &293.0/267.2          &292.8/261.0               &292.1/308.2               &290.8/367.7                  \\
\ce{Xylopyranose <-> AXP_{3-4} + H2O}  &292.9/271.8          &293.2/266.0               &293.5/314.1               &293.1/374.5                  \\
\ce{Xylopyranose <-> AXP_{4-3} + H2O}  &294.5/270.3          &294.2/263.8               &295.8/313.2               &297.0/375.4                  \\
\ce{Xylopyranose <-> AXP_{4-5} + H2O}  &281.6/259.9          &281.1/253.5               &283.3/301.9               &285.6/363.0                  \\ \bottomrule

\end{tabular}%
}
\end{table}
\subsubsection{Thermodynamics of xylose decomposition pathways}
\begin{figure}[htbp]
\centering
\includegraphics[width=0.8\columnwidth]{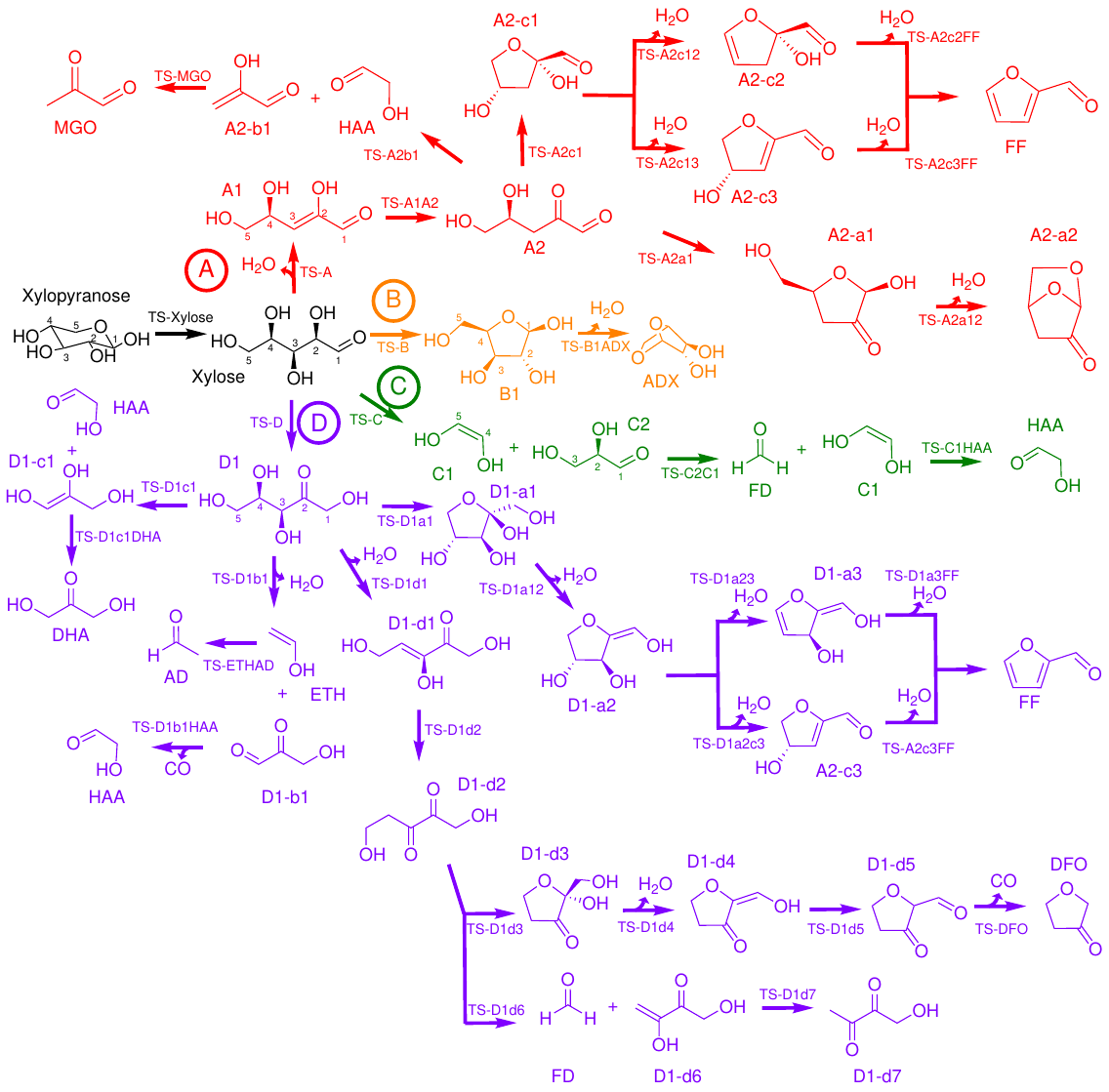} 
\caption{Decomposition pathways of $\beta$-D-xylopyranose. Initial pathway is the ring opening, depicted in black. Subsequent dehydration processes are highlighted in red (A), cyclization in orange (B), C-C bond fission in green (C), and isomerization in purple (D).} 
\label{fig:genscheme}
\end{figure}

\begin{figure}
\centering
\begin{subfigure}[t]{.8\textwidth}
   \includegraphics[width=\linewidth]{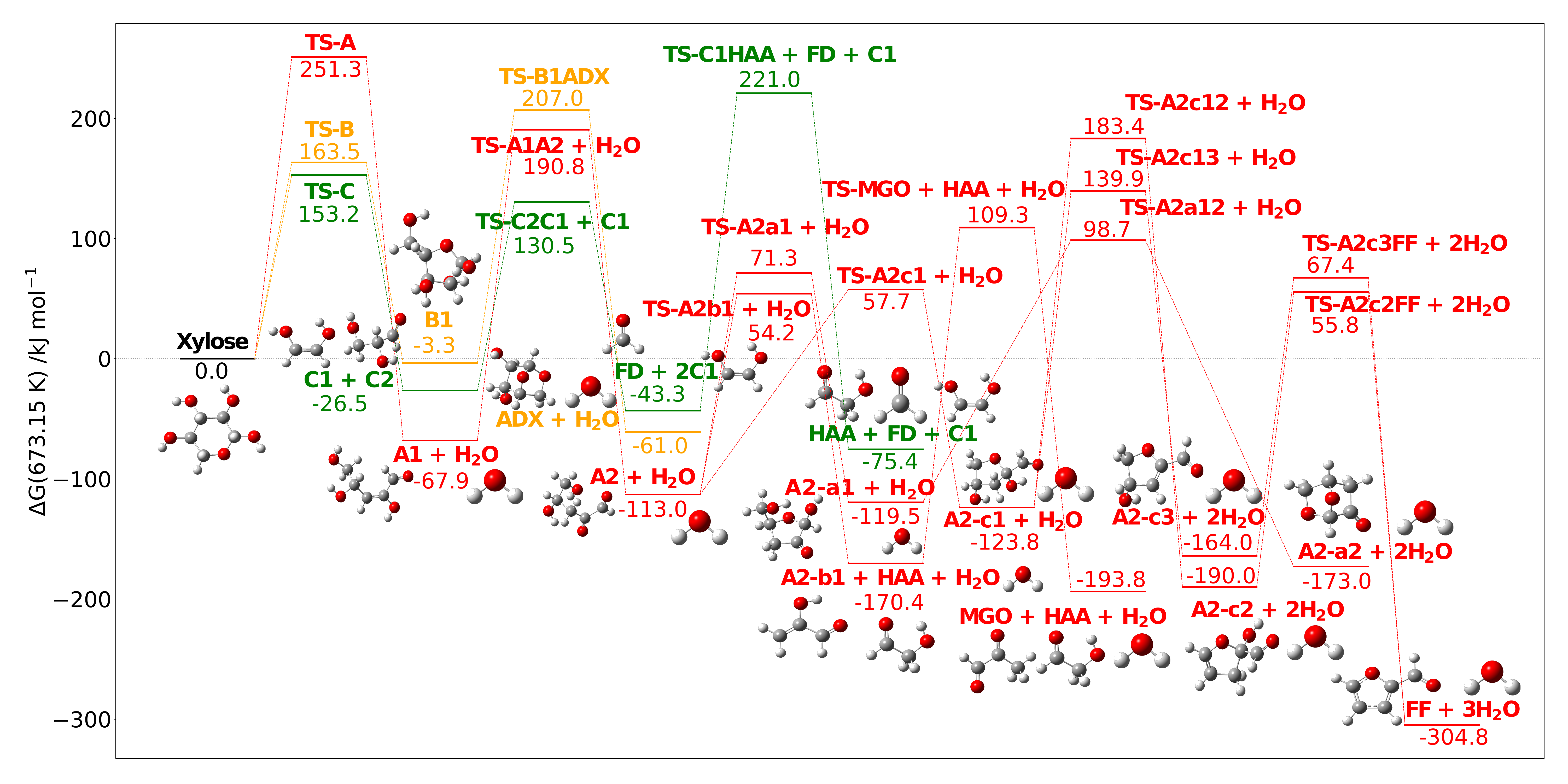}
   \caption{Dehydration, cyclization, and C-C bond fission.}
   \label{fig:pesabc} 
\end{subfigure}
\begin{subfigure}[t]{.8\textwidth}
   \includegraphics[width=\linewidth]{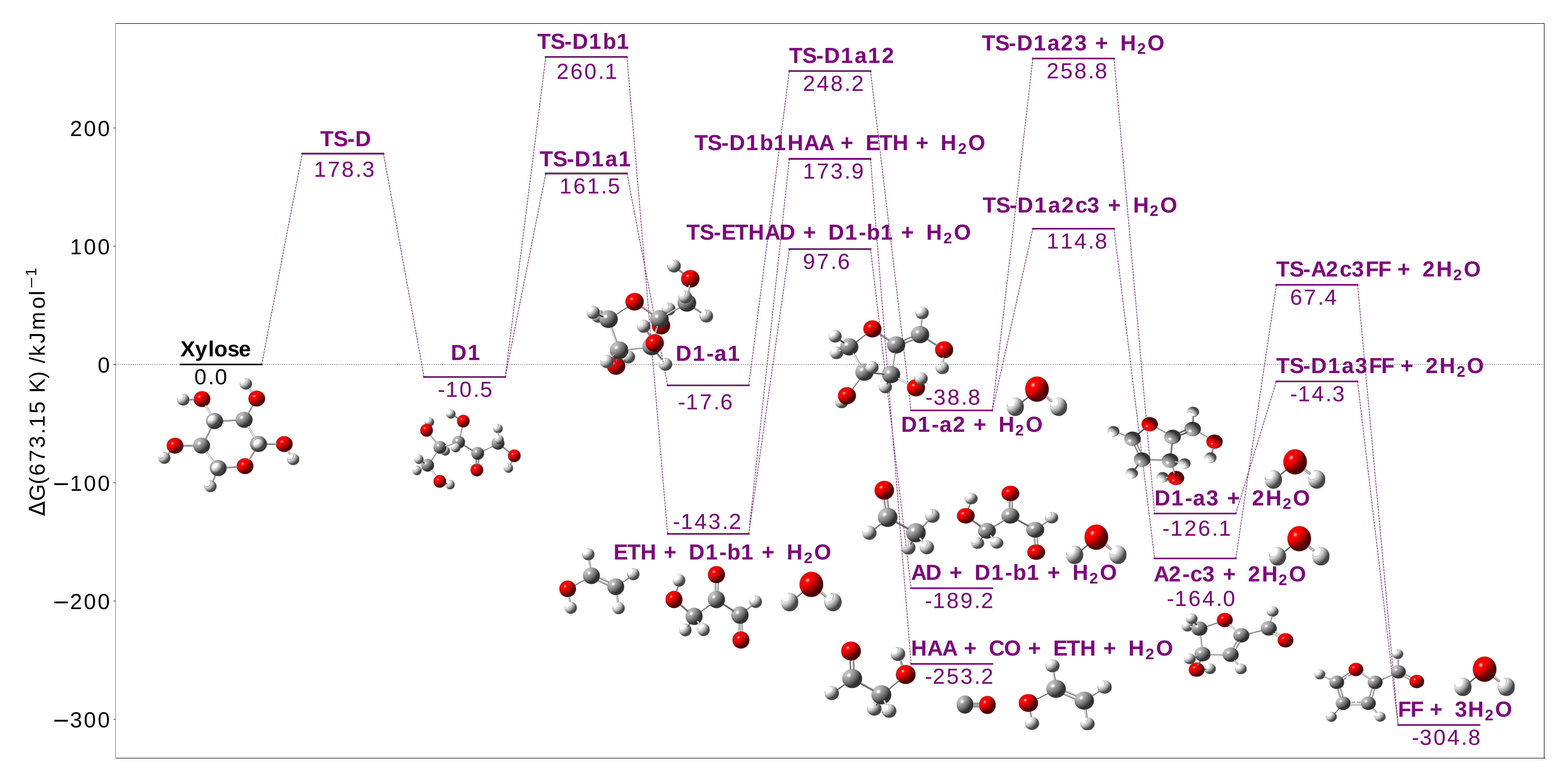}
   \caption{Isomerization part 1.}
   \label{fig:pesd1}
\end{subfigure}
\begin{subfigure}[c]{.80\textwidth}
   \includegraphics[width=\linewidth]{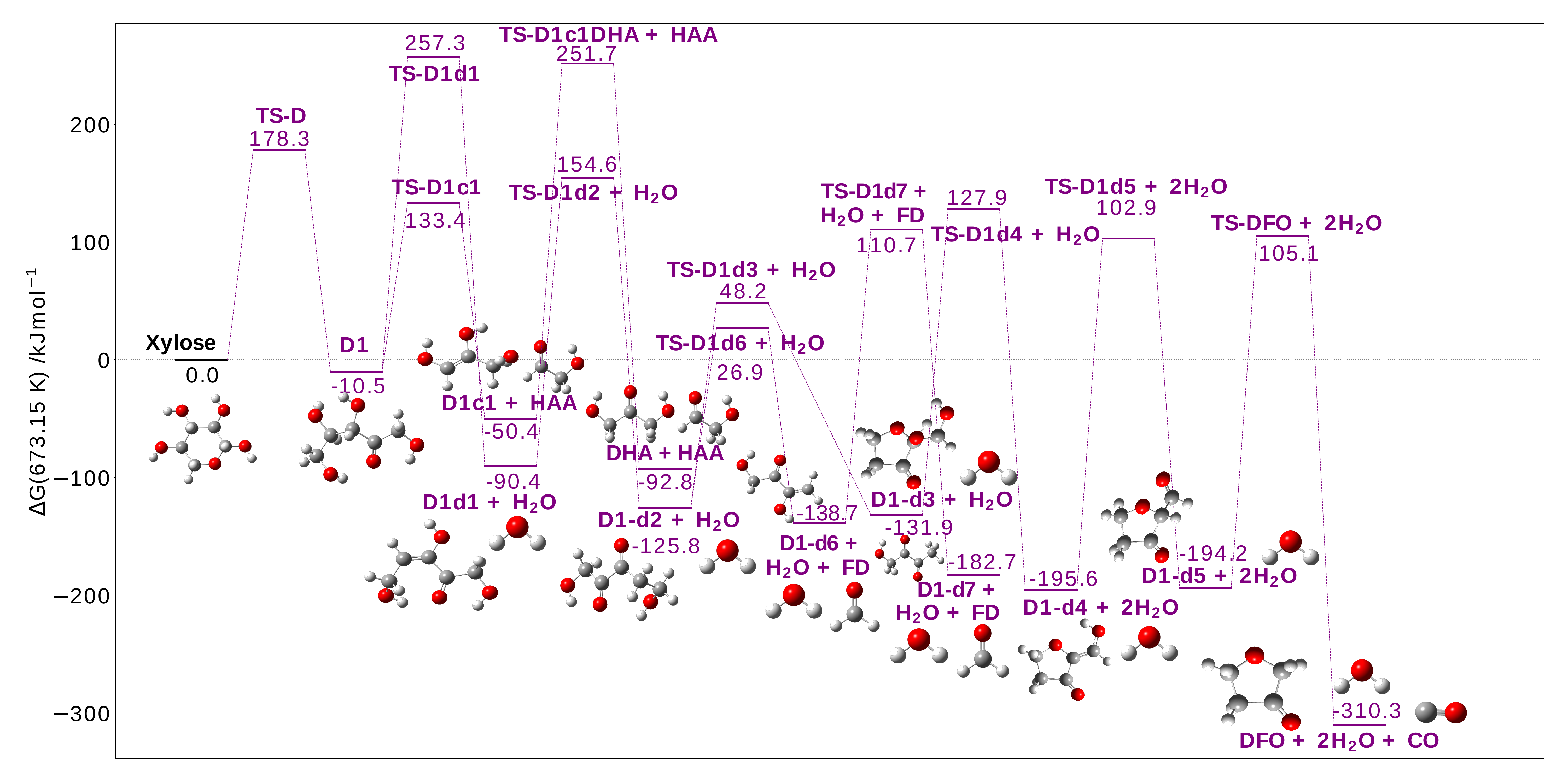}
   \caption{Isomerization part 2.}
   \label{fig:pesd2}
\end{subfigure}
\caption{Relative free energies at \SI{673.15}{\kelvin} of xylose thermal decomposition pathways. Panel (a): dehydration in red, cyclization in orange, C-C bond fission in green. Panels (b), (c): isomerization in purple. $\Delta G$ at DLPNO-CCSD(T)-F12/cc-pVTZ-F12//M06-2X/6-311++G(d,p) level of theory.}
\end{figure}

The upcoming kinetic analysis section will demonstrate that the ring opening pathway is the dominant reaction channel. Consequently, electronic structure calculations were exclusively performed on the reactive pathways associated with the ring opening reaction. \textbf{Xylose} exhibits four distinct types of reactions, depicted in Figure \ref{fig:genscheme}: dehydration (red (A)), cyclization (orange (B)), C-C bond fission (green (C)), and isomerization (purple (D)). The corresponding free energy profiles at \SI{673.15}{\kelvin} are compiled and visualized in Figures \ref{fig:pesabc}, \ref{fig:pesd1}, and \ref{fig:pesd2}. This temperature has been chosen as it mirrors the typical operating temperature during fast pyrolysis.
\paragraph{Dehydration}
Based on the findings presented in Figures \ref{fig:genscheme} and \ref{fig:pesabc}, open-chain D-xylose dehydrates to a trihydroxypentenal (\textbf{A1}) intermediate through a free energy barrier of \SI{251.3}{\kilo\joule\per\mol}. Trihydroxypentenal undergoes tautomerization to its ketoisomer dihydroxyoxopentanal (\textbf{A2}) ($\Delta G^\ddagger_{\SI{673.15}{\kelvin}}=\SI{258.7}{\kilo\joule\per\mol}$). Subsequently, it undergoes decomposition through three distinct pathways, through transition states \textbf{TS-A2a1} ($\Delta G^\ddagger_{\SI{673.15}{\kelvin}}=\SI{ 184.3}{\kilo\joule\per\mol}$), \textbf{TS-A2b1} ($\Delta G^\ddagger_{\SI{673.15}{\kelvin}}=\SI{167.3}{\kilo\joule\per\mol}$), and \textbf{TS-A2c1} ($\Delta G^\ddagger_{\SI{673.15}{\kelvin}}=\SI{170.8}{\kilo\joule\per\mol}$) respectively. Within the first path, dihydroxyoxopentanal undergoes a hemi-acetal reaction between the carbonyl group at the C1 position and the hydroxyl group at the C4 position, resulting in the formation of a five-membered furanone-like intermediate (\textbf{A2-a1}). This intermediate then proceeds to form dioxabicycloheptanone (\textbf{A2-a2}) through an acetal reaction ($\Delta G^\ddagger_{\SI{673.15}{\kelvin}}=\SI{218.2}{\kilo\joule\per\mol}$).
The second path involves the retro-aldol process, which breaks the C3-C4 bond of dihydroxyoxopentanal, resulting in the formation of glycolaldehyde (\textbf{HAA}) and hydroxyacrylaldehyde (\textbf{A2-b1}). Lastly, hydroxyacrylaldehyde undergoes tautomerization ($\Delta G^\ddagger_{\SI{673.15}{\kelvin}}=\SI{279.6}{\kilo\joule\per\mol}$), leading to the formation of methylglyoxal (\textbf{MGO}).
In the third pathway, a hemi-acetal reaction occurs between the carbonyl group at the C2 position and the hydroxyl group at the C5 position, resulting in the formation of a five-membered tetrahydrofurancarbaldehyde-like intermediate (\textbf{A2-c1}). Sequential dehydration reactions take place through \textbf{TS-A2c12} and \textbf{TS-A2c2FF} ($\Delta G^\ddagger_{\SI{673.15}{\kelvin}}=\SI{307.2}{\kilo\joule\per\mol}$ and $\SI{ 245.8}{\kilo\joule\per\mol}$), and \textbf{TS-A2c13} and \textbf{TS-A2c3FF} ($\Delta G^\ddagger_{\SI{673.15}{\kelvin}}=\SI{263.7}{\kilo\joule\per\mol}$ and $\SI{231.4}{\kilo\joule\per\mol}$), leading to the generation of furfural (\textbf{FF}). Out of all the dehydration products, furfural exhibits the highest exoergonicity, with a reaction free energy of \SI{-304.8}{\kilo\joule\per\mol}. In comparison, methylglyoxal possesses a reaction free energy of \SI{-93.8}{\kilo\joule\per\mol}, and dioxabicycloheptanone has a reaction free energy of \SI{-173.0}{\kilo\joule\per\mol}.

\paragraph{Cyclization}
\textbf{Xylose} can undergo either hemi-acetal or acetal reactions during cyclization, with the hemi-acetal reaction being more favorable.\cite{Hu2019} The most favorable pathway is the formation of five-membered intermediate xylofuranose (\textbf{B1}), through transition state \textbf{TS-B} ($\Delta G^\ddagger_{\SI{673.15}{\kelvin}}=\SI{163.5}{\kilo\joule\per\mol}$). Subsequently, xylofuranose undergoes an acetal reaction to form the 1,5-acetal ring, resulting in the formation of anhydro-D-xylopyranose (\textbf{ADX}), which lies at \SI{-61.0}{\kilo\joule\per\mol}. This step has the highest free energy barrier of this pathway ($\Delta G^\ddagger_{\SI{673.15}{\kelvin}}=\SI{210.2}{\kilo\joule\per\mol}$).
\paragraph{Carbon-carbon bond scission}
According to Hu \emph{et al.},\cite{Hu2019} the degradation of \textbf{xylose} involves two primary mechanisms: retro-aldol reaction and cyclic Grob fragmentation, both with relatively low activation energies. Among these mechanisms, the retro-aldol reaction is found to be the most favorable. Hence, the study focuses solely on this reaction. Initially, \textbf{xylose} undergoes a retro-aldol reaction, leading to the formation of intermediate glyceraldehyde (\textbf{C2}) and the enol isomer of glycolaldehyde (\textbf{C1}) through the cleavage of the C2-C3 bond ($\Delta G^\ddagger_{\SI{673.15}{\kelvin}} = \SI{153.2}{\kilo\joule\per\mol}$). Subsequently, glyceraldehyde undergoes another retro-aldol reaction to generate the enol isomer ($\Delta G^\ddagger_{\SI{673.15}{\kelvin}} = \SI{157.0}{\kilo\joule\per\mol}$). Lastly, the enol undergoes an enol-keto tautomerization process, resulting in the formation of glycolaldehyde (\textbf{HAA}). This particular step has the higher free energy barrier on the sequence for glycolaldehyde formation, so will impact its rate of formation most significantly ($\Delta G^\ddagger_{\SI{673.15}{\kelvin}} = \SI{264.2}{\kilo\joule\per\mol}$).
\paragraph{Isomerization}

\textbf{Xylose} can undergo an isomerization reaction. As an open-chain aldose, it readily undergoes isomerization to its ketone isomer, D-xylulose (\textbf{D1}), with an activation free energy of \SI{178.3}{\kilo\joule\per\mol}. Similar to \textbf{xylose}, D-xylulose undergoes decomposition through cyclization, C-C bond cleavage, and dehydration reactions. Through transition state \textbf{TS-D1a1} ($\Delta G^\ddagger_{\SI{673.15}{\kelvin}} = \SI{171.9}{\kilo\joule\per\mol}$) D-xylulose forms a five-membered tetrahydrofuran-like intermediate (\textbf{D1-a1}), through a hemi-acetal reaction. Subsequently, it undergoes dehydration at the 2-OH\&1-H sites, generating \textbf{D1-a2} ($\Delta G^\ddagger_{\SI{673.15}{\kelvin}} = \SI{265.8}{\kilo\joule\per\mol}$). This intermediate further produces furfural (\textbf{FF}) through successive dehydration reactions. The dehydration at the 2-OH\&1-H sites (\textbf{D1-a1} to \textbf{D1-a2}) serves as the rate step for furfural formation. In contrast, the dehydration at the 4-OH\&5-H sites (\textbf{D1-a2} to \textbf{D1-a3}) may limit the rate of furfural formation in the concurrent path, as the activation free energy of $\Delta G^\ddagger_{\SI{673.15}{\kelvin}} = \SI{297.6}{\kilo\joule\per\mol}$ is \SI{32}{\kilo\joule\per\mol} higher. Moreover, D-xylulose undergoes cyclic Grob fragmentation at the 5-OH\&3-H sites, generating ethenol (\textbf{ETH}) and 3-hydroxy-2-oxopropanal (\textbf{D1-b1}) with an activation free energy of \SI{270.5}{\kilo\joule\per\mol}. Ethenol then tautomerizes into acetaldehyde (\textbf{AD}), with an activation free energy of \SI{240.9}{\kilo\joule\per\mol} above reactant. 3-hydroxy-2-oxopropanal undergoes decarbonylation to form glycolaldehyde (\textbf{HAA}) ($\Delta G^\ddagger_{\SI{673.15}{\kelvin}}=\SI{317.2}{\kilo\joule\per\mol}$). The C3-C4 bond of D-xylulose undergoes cleavage via a retro-aldol reaction, resulting in the formation of glycolaldehyde (\textbf{HAA}) and \textbf{D1-c1} $\Delta G^\ddagger_{\SI{673.15}{\kelvin}} = \SI{143.8}{\kilo\joule\per\mol}$). It then generates 1,3-dihydroxyacetone (\textbf{DHA}) through tautomerization. The reaction step with the highest $\Delta G^\ddagger$ barrier for 1,3-dihydroxyacetone formation is the tautomerization of \textbf{D1-c1} into itself, with a \SI{302.1}{\kilo\joule\per\mol} free energy barrier.
In addition to the previously mentioned pathways, the dehydration at 4-OH\&3-H sites ($\Delta G^\ddagger_{\SI{673.15}{\kelvin}}=\SI{267.8}{\kilo\joule\per\mol}$) and following tautomerization ($\Delta G^\ddagger_{\SI{673.15}{\kelvin}}=\SI{244.9}{\kilo\joule\per\mol}$) result in the formation of dihydroxypentanedione (\textbf{D1-d2}) which decomposes through TSs \textbf{TS-D1d3} and \textbf{TS-D1d6}, with activation energies of \SI{174.1}{\kilo\joule\per\mol} and \SI{152.7}{\kilo\joule\per\mol}, respectively. In the former case dihydroxypentanedione undergoes a hemi-acetal reaction, forming the five-membered  furanone-like intermediate (\textbf{D1-d3}). Subsequently, it undergoes successive dehydration reactions at the 2-OH\&1-H sites, followed by tautomerization and decarbonylation, with activation energies, in order, of 259.8, 298.5, and \SI{299.3}{\kilo\joule\per\mol} leading to the formation of dihydrofuran-3(2H)-one (\textbf{DFO}). The highest free energy barrier reaction step in this path is the tautomerization of dehydrated furanone-like intermediate \textbf{D1-d4}. In the latter case, dihydroxypentanedione undergoes a retro-aldol reaction, resulting in the cleavage of the C5-C6 bond and the formation of formaldehyde (\textbf{FD}) and dihydroxybutenone (\textbf{D1-d6}), with an activation energy of \SI{152.7}{\kilo\joule\per\mol}. Lastly, dihydroxybutenone tautomerizes into hydroxybutanedione (\textbf{D1-d7}) overcoming a free energy barrier of \SI{249.4}{\kilo\joule\per\mol}. Formation of dihydrofuran-3(2H)-one is the most exergonic process with a relative free energy of \SI{-310.3}{\kilo\joule\per\mol}, followed by furfural \SI{-304.8}{\kilo\joule\per\mol}.

Detailed forward and reverse reaction barriers at \SI{0}{\kelvin}, standard, and operative conditions are reported in Table \ref{tab:barriersdxylose}. Enthalpies of formation, entropies and heat capacities at constant pressure at different temperatures are reported in Table \ref{tab:therm}. 
As described in the Computational Methodology section, enthalpies of formation were determined by using the the enthalpies of atomization. To further validate our findings, in Table \ref{tab:therm} are included the enthalpy of formation values from ATcT for commonly known species: \ce{H2O} (\SI{-241.80}{\kilo\joule\per\mol}), ethenol (ETH) (\SI{-123.72}{\kilo\joule\per\mol}), acetaldehyde (AD) (\SI{-165.55}{\kilo\joule\per\mol}), CO (\SI{-110.52}{\kilo\joule\per\mol}), formaldehyde (FD) (\SI{-109.23}{\kilo\joule\per\mol}), and hydroxyacetaldehyde (\SI{-317.50}{\kilo\joule\per\mol}). The errors in our methodology are quantified by a mean unsigned error (MUE) of \SI{6.35}{\kilo\joule\per\mol} and a root mean square deviation (RMSD) of \SI{7.50}{\kilo\joule\per\mol}. It should be noted that this paper does not focus on benchmarking the enthalpy of formation; rather, the provided numbers serve as reference points to gauge the effectiveness of the employed approach. For greater accuracy, methods that incorporate error cancellation, such as the isodesmic approach, should be considered.
As far as the authors know, this is the first instance in the literature where thermochemical parameters for the pyrolytic reaction system of xylopyranose have been directly derived from first principles.
\begin{table}[htbp]
\caption{Activation enthalpies and free energies at \SI{0}{\kelvin}, \SI{298.15}{\kelvin}, and \SI{673.15}{\kelvin} at DLPNO-CCSD(T)-F12/cc-pVTZ-F12//M06-2X/6-311++G(d,p) for the single-step reactions of D-xylose thermal decomposition pathways. Pressure is set at 1 atm. Energies in \SI{}{\kilo\joule\per\mol}.}
\label{tab:barriersdxylose}
\begin{tabular}{@{}lcccc@{}}
\toprule
Reaction         & \multicolumn{1}{c}{$\Delta H^\ddagger_{\SI{0}{\kelvin}}$} &$\Delta H^\ddagger_{\SI{298.15}{\kelvin}}$ &$\Delta G^\ddagger_{\SI{298.15}{\kelvin}}$ &$\Delta G^\ddagger_{\SI{673.15}{\kelvin}}$ \\ \midrule
                                   &\multicolumn{4}{c}{forward/reverse}            \\\midrule
\ce{Xylose <-> A{1} + H2O}              &263.8/213.5     &265.4/206.5       &259.4/256.1     &251.3/319.2       \\
\ce{A{1} <-> A{2}}                 &258.9/310.2     &258.1/310.2       &257.8/307.2     &258.7/303.8       \\
\ce{A{2} <-> A{2}-a{1}}            &172.9/195.0     &170.4/195.3       &175.9/193.1     &184.3/190.8       \\
\ce{A{2}-a{1} <-> A{2}-a{2} + H2O} &215.3/178.8     &214.9/173.8       &216.3/217.3     &218.2/271.8       \\
\ce{A{2} <-> A{2}-b{1} + HAA}      &164.7/101.9     &163.4/96.3        &164.7/152.6     &167.3/224.6       \\
\ce{A{2}-b{1} <-> MGO}             &282.6/297.2     &282.4/296.3       &280.8/299.2     &279.6/303.1       \\
\ce{A{2} <-> A{2}-c{1}}            &158.6/183.1     &156.5/183.5       &162.4/182.5     &170.8/181.5       \\
\ce{A{2}-c{1} <-> A{2}-c{2} + H2O} &312.6/275.1     &314.0/269.6       &311.6/315.7     &307.2/373.4       \\
\ce{A{2}-c{2} <-> FF + H2O}        &250.4/265.0     &252.1/260.1       &249.8/304.9     &245.8/360.6       \\
\ce{A{2}-c{1} <-> A{2}-c{3} + H2O} &268.7/201.2     &269.6/194.5       &267.1/242.6     &263.7/303.9       \\
\ce{A{2}-c{3} <-> FF + H2O}        &234.0/278.6     &234.8/273.5       &233.6/317.3     &231.4/372.2       \\
\ce{Xylose <-> B{1}}                  &157.3/163.5     &156.3/163.1       &159.1/164.4     &163.5/166.7       \\
\ce{B{1} <-> ADX + H2O}            &208.5/178.9     &208.2/174.6       &208.9/216.0     &210.2/267.9       \\
\ce{Xylose <-> C{1} + C{2}}           &159.0/43.6      &159.8/38.1        &156.8/100.6     &153.2/179.7       \\
\ce{C{2} <-> FD + C{1}}            &147.1/61.5      &146.4/55.7        &151.1/107.9     &157.0/173.8       \\
\ce{C{1} <-> HAA}                  &261.0/294.0     &260.6/293.8       &261.9/295.0     &264.2/296.4       \\
\ce{Xylose <-> D{1}}                  &179.6/176.2     &178.6/173.4       &177.9/179.4     &178.3/188.7       \\
\ce{D{1} <-> D{1}-a{1}}            &155.1/182.6     &153.0/183.0       &161.0/181.0     &171.9/179.1       \\
\ce{D{1}-a{1} <-> D{1}-a{2} + H2O} &277.5/191.0     &279.3/185.6       &273.5/230.5     &265.8/287.1       \\
\ce{D{1}-a{2} <-> D{1}-a{3} + H2O} &305.1/292.5     &306.3/287.7       &303.0/330.8     &297.6/384.8       \\
\ce{D{1}-a{3} <-> FF + H2O}        &104.8/182.3     &103.8/175.0       &107.0/225.9     &111.8/290.5       \\
\ce{D{1}-a{2} <-> A{2}-c{3} + H2O} &148.7/169.0     &147.9/161.8       &150.2/213.2     &153.6/278.8       \\
\ce{D{1} <-> D{1}-b{1} + ETH + H2O}&268.5/177.5     &267.9/166.1       &269.1/271.1     &270.5/403.3       \\
\ce{ETH <-> AD}                    &238.0/280.4     &237.3/279.2       &238.5/282.4     &240.9/286.8       \\
\ce{D{1}-b{1} <-> HAA + CO}        &330.7/346.5     &331.8/342.9       &325.4/380.2     &317.2/427.2       \\
\ce{D{1} <-> D{1}-c{1} + HAA}      &135.4/54.0      &134.4/48.8        &138.4/108.3     &143.8/183.7       \\
\ce{D{1}-c{1} <-> DHA}             &306.0/345.0     &306.2/345.6       &304.0/345.2     &302.1/344.6       \\ 
\ce{D{1} <-> D{1}-d{1} + H2O}      &269.3/248.7     &269.3/242.2       &268.8/288.6     &267.8/347.7       \\
\ce{D{1}-d{1} <-> D{1}-d{2}}       &246.6/277.0     &246.1/276.9       &245.1/278.5     &244.9/280.4       \\
\ce{D{1}-d{2} <-> D{1}-d{3}}       &152.7/178.1     &150.2/177.9       &160.4/178.7     &174.1/180.2       \\
\ce{D{1}-d{3} <-> D{1}-d{4} + H2O} &267.8/228.1     &270.5/223.9       &266.4/268.3     &259.8/323.5       \\
\ce{D{1}-d{4} <-> D{1}-d{5}}       &298.8/298.1     &298.3/298.1       &298.0/297.5     &298.5/297.1       \\
\ce{D{1}-d{5} <-> DFO + CO}        &308.9/325.2     &310.2/321.9       &305.6/363.6     &299.3/415.4       \\
\ce{D{1}-d{2} <-> FD + D{1}-d{6}}  &143.9/51.8      &142.4/45.2        &146.8/98.1      &152.7/165.6       \\
\ce{D{1}-d{6} <-> D{1}-d{7}}       &250.4/287.4     &250.0/286.7       &249.2/289.6     &249.4/293.3       \\ \bottomrule
\end{tabular}
\end{table}

\begin{table}[htbp]
\caption{Enthalpies of formation ($\Delta H_f^\standardstate $) and entropies ($S^\standardstate$) of molecules at \SI{298.15}{\kelvin} and 1 atm, and heat capacities ($C_p$) at different temperatures. In parentheses, values taken from ATcT. Nomenclature is the same used in Figures \ref{fig:initialsteps} and \ref{fig:genscheme}. Units are \SI{}{\kilo\joule\per\mol} for $\Delta H_f^\standardstate$ and \SI{}{\joule\per\mol\per\kelvin} for $S^\standardstate$ and $C_p$.}
\label{tab:therm}
\resizebox{\textwidth}{!}{%
\begin{tabular}{@{}lcccccccccc@{}}
\toprule
     &  $\Delta H_f^\standardstate $       &  $S^\standardstate$      & $C_p(\SI{300}{\kelvin})$    & $C_p(\SI{400}{\kelvin})$    & $C_p(\SI{500}{\kelvin})$    & $C_p(\SI{600}{\kelvin})$    & $C_p(\SI{800}{\kelvin})$    & $C_p(\SI{1000}{\kelvin})$   & $C_p(\SI{1500}{\kelvin})$ & Chemical Formula   \\ \midrule
Xylopyranose & -881.85 & 400.15 & 168.60 & 212.33 & 250.57 & 281.98 & 328.47 & 360.92 & 410.09 &\ce{C5H10O5} \\
Xylose      & -863.55 & 409.57 & 171.62 & 215.28 & 252.83 & 283.55 & 329.18 & 361.29 & 410.35 &\ce{C5H10O5} \\
AXP$_{1}$   & -579.18 & 374.16 & 145.63 & 183.44 & 215.83 & 242.12 & 280.73 & 307.53 & 348.08 &\ce{C5H8O4} \\
AXP$_{2-1}$ & -619.09 & 378.62 & 145.71 & 183.62 & 216.06 & 242.34 & 280.86 & 307.59 & 348.05 &\ce{C5H8O4} \\
AXP$_{2-3}$ & -609.46 & 372.30 & 142.96 & 181.49 & 214.35 & 240.96 & 279.99 & 307.05 & 347.90 &\ce{C5H8O4} \\
AXP$_{3-2}$ & -606.71 & 372.20 & 143.30 & 181.82 & 214.64 & 241.21 & 280.16 & 307.17 & 347.94 &\ce{C5H8O4} \\
AXP$_{3-4}$ & -611.31 & 371.74 & 144.00 & 182.39 & 215.10 & 241.58 & 280.41 & 307.34 & 348.00 &\ce{C5H8O4} \\
AXP$_{4-3}$ & -608.00 & 372.20 & 144.38 & 182.17 & 214.65 & 241.12 & 280.16 & 307.30 & 348.21 &\ce{C5H8O4} \\
AXP$_{4-5}$ & -610.90 & 366.35 & 141.94 & 181.27 & 214.60 & 241.40 & 280.37 & 307.26 & 347.86 &\ce{C5H8O4} \\
TH-C-OH2-c  & -593.26 & 379.99 & 141.11 & 178.95 & 212.07 & 239.24 & 279.45 & 307.41 & 349.06 &\ce{C5H8O4} \\
TH-C-OH2-t  & -590.67 & 381.12 & 142.38 & 180.11 & 213.09 & 240.09 & 279.98 & 307.70 & 349.10 &\ce{C5H8O4} \\
\ce{H2O}  & -243.39 & 188.58 & 33.49  & 34.14  & 35.09  & 36.14  & 38.31  & 40.56  & 45.79  &\ce{H2O} \\
     & (-241.80) &        &        &        &        &        &        &        &        &             \\   
A1   & -561.33 & 407.14 & 156.54 & 191.10 & 220.99 & 245.64 & 282.60 & 308.75 & 348.74 &\ce{C5H8O4} \\
A2   & -613.36 & 398.08 & 150.92 & 185.97 & 216.65 & 242.12 & 280.54 & 307.77 & 348.99 &\ce{C5H8O4} \\
A2-a1 & -638.29 & 372.11 & 138.94 & 177.85 & 211.35 & 238.63 & 278.87 & 306.83 & 348.61 &\ce{C5H8O4} \\
A2-a2 & -353.77 & 324.63 & 102.53 & 137.73 & 168.03 & 192.47 & 227.99 & 252.14 & 286.96 &\ce{C5H6O3} \\
A2-b1 & -239.54 & 297.75 & 80.64  & 98.67  & 113.78 & 126.12 & 144.71 & 158.00 & 178.41 &\ce{C3H4O2} \\
A2-c1 & -640.37 & 374.94 & 141.34 & 179.95 & 213.12 & 240.11 & 279.93 & 307.62 & 349.07 &\ce{C5H8O4} \\
A2-c2 & -352.54 & 349.10 & 119.23 & 151.44 & 178.62 & 200.53 & 232.59 & 254.72 & 287.45 &\ce{C5H6O3} \\
A2-c3 & -321.83 & 356.25 & 119.05 & 149.98 & 176.78 & 198.72 & 231.19 & 253.69 & 286.94 &\ce{C5H6O3} \\
B1   & -870.30 & 404.95 & 167.49 & 212.32 & 251.14 & 282.70 & 328.99 & 361.15 & 410.02 &\ce{C5H10O5} \\
C1   & -273.58 & 285.09 & 71.62  & 86.89  & 99.80  & 110.21 & 125.62 & 136.66 & 154.38 &\ce{C2H4O2} \\
C2   & -468.24 & 344.19 & 107.31 & 129.65 & 149.81 & 166.81 & 192.78 & 211.45 & 240.27 &\ce{C3H6O3} \\
D1   & -858.35 & 432.34 & 179.13 & 220.24 & 256.24 & 286.00 & 330.57 & 362.10 & 410.53 &\ce{C5H10O5} \\
D1-a1 & -888.39 & 398.66 & 167.16 & 212.91 & 251.91 & 283.39 & 329.38 & 361.32 & 410.01 &\ce{C5H10O5} \\
D1-a2 & -551.37 & 379.80 & 146.00 & 183.41 & 215.63 & 241.90 & 280.61 & 307.54 & 348.22 &\ce{C5H8O4} \\
D1-a3 & -289.31 & 347.18 & 120.49 & 153.03 & 180.15 & 201.76 & 232.99 & 254.39 & 286.41 &\ce{C5H6O3} \\
D1-b1 & -393.44 & 335.53 & 96.73  & 114.71 & 130.71 & 144.24 & 165.02 & 179.84 & 202.08 &\ce{C3H4O3} \\
D1-c1 & -465.91 & 334.00 & 107.56 & 131.65 & 152.17 & 168.90 & 193.79 & 211.51 & 239.37 &\ce{C3H6O3} \\
D1-d1 & -587.81 & 401.20 & 155.69 & 190.43 & 220.44 & 245.17 & 282.25 & 308.48 & 348.61 &\ce{C5H8O4} \\
D1-d2 & -618.65 & 409.75 & 153.28 & 187.19 & 217.22 & 242.34 & 280.49 & 307.65 & 348.88 &\ce{C5H8O4} \\
D1-d3 & -646.38 & 378.07 & 142.23 & 180.31 & 213.11 & 239.87 & 279.48 & 307.12 & 348.63 &\ce{C5H8O4} \\
D1-d4 & -356.41 & 352.05 & 117.40 & 148.64 & 175.56 & 197.60 & 230.26 & 252.94 & 286.50 &\ce{C5H6O3} \\
D1-d5 & -356.18 & 351.27 & 114.60 & 145.85 & 173.19 & 195.78 & 229.59 & 253.16 & 287.56 &\ce{C5H6O3} \\
D1-d6 & -417.52 & 353.82 & 120.65 & 147.04 & 169.31 & 187.50 & 214.69 & 234.02 & 263.77 &\ce{C4H6O3} \\
D1-d7 & -454.25 & 365.85 & 120.52 & 144.80 & 166.40 & 184.66 & 212.84 & 233.17 & 264.16 &\ce{C4H6O3} \\
DFO  & -270.05 & 309.03 & 87.36  & 114.04 & 137.85 & 157.71 & 187.74 & 208.97 & 240.49 &\ce{C4H6O2} \\
ETH  & -119.77 & 256.18 & 56.08  & 69.14  & 80.44  & 89.76  & 103.98 & 114.48 & 131.47 &\ce{C2H4O} \\
     & (-123.72) &        &        &        &        &        &        &        &        &             \\  
FF   & -117.08 & 318.60 & 91.29  & 116.89 & 138.81 & 156.52 & 182.29 & 199.77 & 224.85 &\ce{C5H4O2} \\
MGO  & -253.47 & 312.78 & 82.58  & 98.26  & 112.38 & 124.53 & 143.70 & 157.73 & 179.02 &\ce{C3H4O2} \\
DHA  & -505.26 & 340.30 & 105.62 & 128.26 & 148.65 & 165.85 & 192.19 & 211.14 & 240.30 &\ce{C3H6O3} \\
AD   & -161.67 & 262.97 & 55.20  & 66.29  & 76.99  & 86.51  & 102.04 & 113.79 & 132.21 &\ce{C2H4O} \\
     & (-165.55) &        &        &        &        &        &        &        &        &             \\     
\ce{CO}   & -97.79  & 197.40 & 29.12  & 29.25  & 29.60  & 30.15  & 31.48  & 32.71  & 34.75 &\ce{CO}  \\
          & (-110.52) &        &        &        &        &        &        &        &        &             \\   
ADX  & -593.37 & 352.68 & 129.44 & 171.17 & 207.15 & 236.09 & 277.89 & 306.35 & 348.36 &\ce{C5H4O4} \\
FD   & -103.99 & 218.48 & 35.14  & 38.61  & 42.84  & 47.10  & 54.66  & 60.66  & 70.15  &\ce{CH2O} \\
     & (-109.23) &        &        &        &        &        &        &        &        &          \\
HAA  & -306.78 & 284.71 & 67.95  & 82.06  & 95.00  & 106.08 & 123.33 & 135.90 & 155.27 &\ce{C2H4O2} \\
     & (-317.50) &        &        &        &        &        &        &        &        &            \\
\bottomrule
\end{tabular}
}
\end{table}
A final remark is deserved. The role of free energies, as opposed to enthalpies, has been emphasized in this section, with a particular focus on the significance of the Gibbs energy of activation in the examination of mechanisms and kinetics in chemical reactions. This emphasis is attributed to the fact that it accounts for both enthalpic and entropic corrections associated with electronic and zero-point energy components. It is essential to recognize that, especially at low temperatures or energies, certain reaction pathways may initially appear less favorable primarily due to their enthalpy contributions. However, as temperatures and energies increase, entropic factors begin to exert their influence, potentially making these pathways more dominant. Therefore, the prediction and exploration of activation free energy become of paramount importance when the primary goal is to gain insights into reaction kinetics and to discern the prevailing mechanistic pathways. This approach offers a more comprehensive and enlightening perspective in the field of chemical reaction studies.
\subsection{Reaction kinetics}
\begin{figure}[htbp]
\centering
\subfloat[][Xylopyranose initial thermal decomposition.]
{\includegraphics[width=0.5\textwidth]{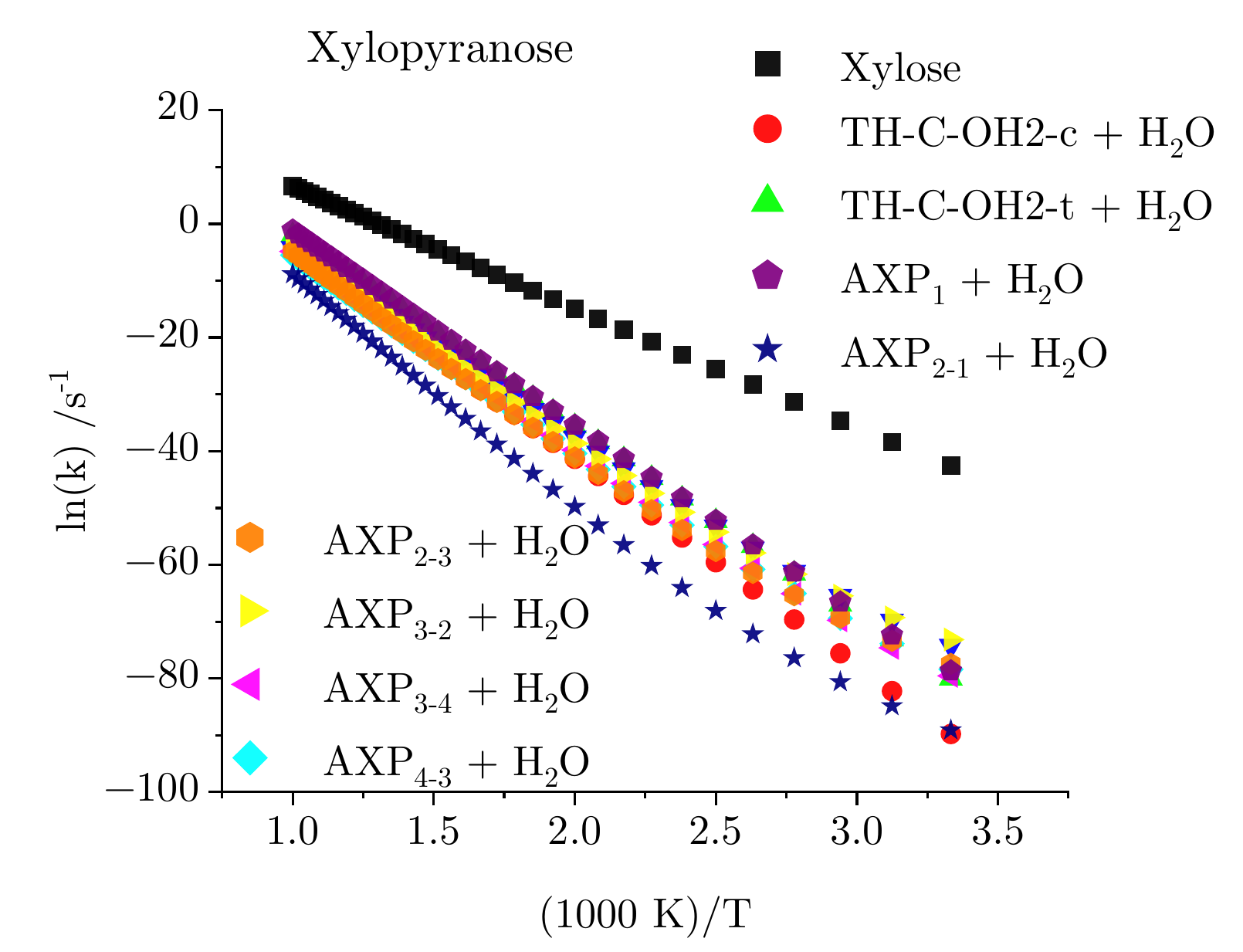}}
\subfloat[][Xylose thermal decomposition.]
{\includegraphics[width=0.5\textwidth]{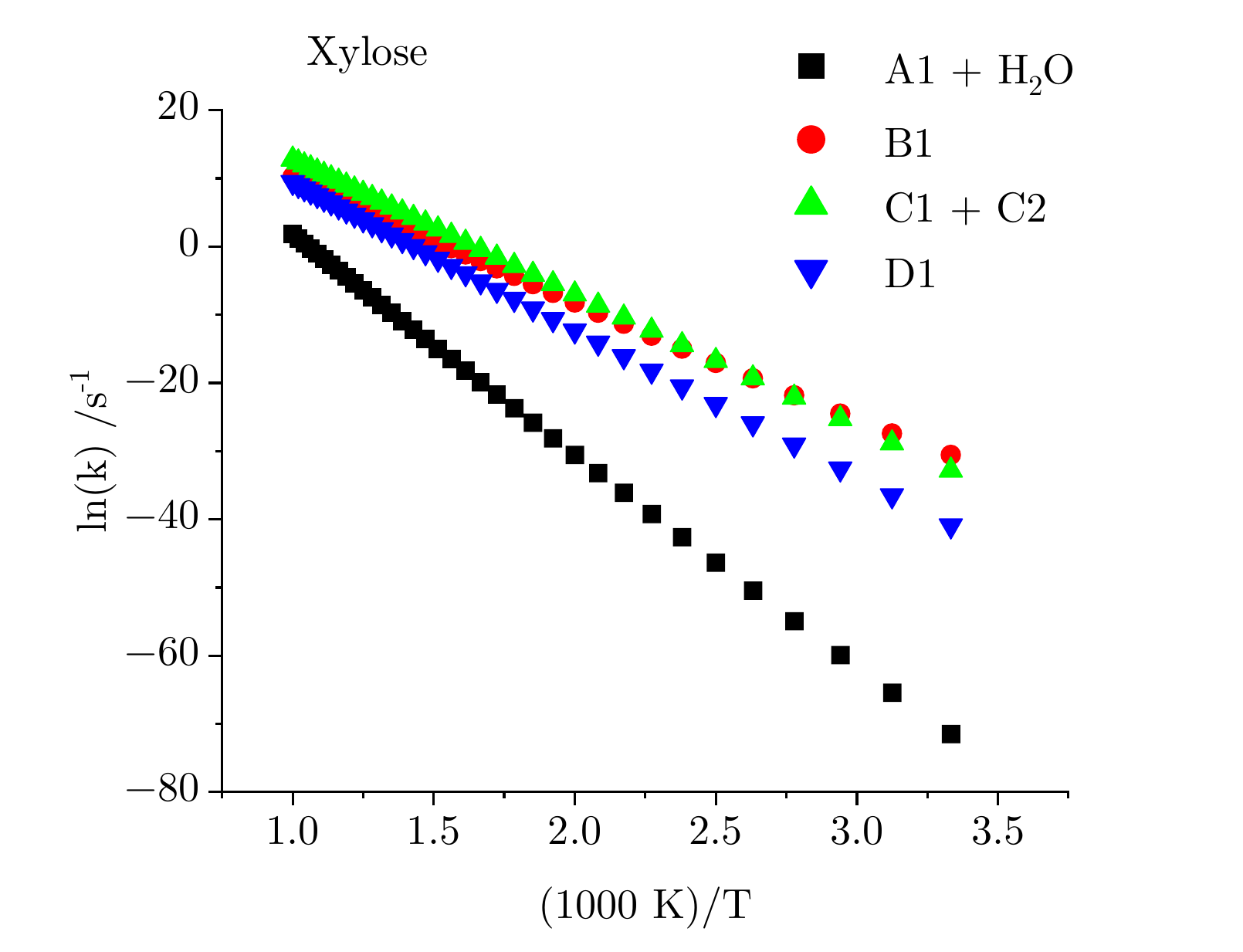}} \qquad
\caption{Temperature dependent rate constants for xylopyranose decomposition (panel (a), xylose ring opening, TH-C-OH2-(c,t) ring contraction, AXP$_\mathrm{n}$ dehydration) and D-xylose decomposition (panel (b), \ce{A{1} + H2O} dehydration, B1 cyclization, \ce{C{1} + C{2}} carbon-carbon bond fission, and D1 isomerization).}
\label{fig:rates_xylo}
\end{figure}

The determination of temperature-dependent reaction rate constants for the initial decomposition pathways of \textbf{xylopyranose} reveals that the ring opening channel dominates across the entire temperature range (\num{300}-\SI{1000}{\kelvin}), as shown in Figure \ref{fig:rates_xylo} (a). Figure \ref{fig:rates_xylo} (b) shows the temperature-dependent rate coefficients of \textbf{xylose} decomposition pathways. At very high temperatures, there is a strong competition between all four reaction pathways, with the C-C bond scission leading to \textbf{C1 + C2} pathway proceeding at the fastest rate. However, as the temperature decreases the isomerization channel leading to \textbf{D1} and the dehydration pathway leading to \textbf{\ce{A{1} + H2O}} become notably slower. Furthermore, as the temperature decreases below \SI{400}{\kelvin}, the cyclization channel leading to \textbf{B1} becomes the most reactive pathway. Notably, each reaction rate constant appears to obey an Arrhenius-like temperature dependence. 

\begin{table}[htbp]
\caption{Arrhenius parameters obtained through fitting of rate constants in the \num{300}-\SI{1000}{\kelvin} temperature range for the initial thermal decomposition pathways of xylopyranose and subsequent pathways originating from the open-chain form of D-xylose.}
\label{tab:arrhpar}
\begin{tabular}{@{}lccc@{}}
\toprule
Reaction       & $A$ /\SI{}{\per\second}        & $E_a$ /\SI{}{\kilo\joule\per\mol} & rss \textsuperscript{a}   \\ \midrule
\ce{Xylopyranose -> Xylose}                & \num{2.53E+12} & 181.8 & \num{7.3d-2}\\
\ce{Xylopyranose -> TH-C-OH{2}-c + H2O}      & \num{7.08E+13} & 305.1 & \num{1.8d-13} \\
\ce{Xylopyranose -> TH-C-OH{2}-t + H2O}      & \num{6.42E+13} & 280.8 & \num{7.9d-11} \\
\ce{Xylopyranose -> AXP1 + H2O}         & \num{4.60E+14} & 289.7 & \num{4.7d-9}\\
\ce{Xylopyranose -> AXP_{2-1} + H2O}    & \num{2.76E+14} & 349.9 & \num{9.9d-16}\\
\ce{Xylopyranose -> AXP_{2-3} + H2O}    & \num{1.23E+14} & 309.8 & \num{4.6d-12}\\
\ce{Xylopyranose -> AXP_{3-2} + H2O}    & \num{4.58E+13} & 295.4 & \num{2.6d-11}\\
\ce{Xylopyranose -> AXP_{3-4} + H2O}    & \num{2.32E+13} & 296.2 & \num{3.6d-12}\\
\ce{Xylopyranose -> AXP_{4-3} + H2O}    & \num{1.26E+13} & 296.5 & \num{1.1d-12}\\
\ce{Xylopyranose -> AXP_{4-5} + H2O}    & \num{8.83E+12} & 283.3 & \num{1.6d-11} \\
\ce{Xylose -> A{1} + H2O}        & \num{1.17E+15} & 273.4 & \num{7.3d-8}      \\
\ce{Xylose -> B{1}}              & \num{4.17E+12} & 156.4 & \num{12.4}     \\
\ce{Xylose -> C{1} + C{2}}       & \num{1.76E+14} & 166.5 & \num{421.5}     \\
\ce{Xylose -> D{1}}              & \num{4.38E+13} & 182.7 & \num{6d-1}     \\ \bottomrule
\end{tabular}\\
\textsuperscript{a}rss stands for residual sum of squares of the fit.
\end{table}
\subsection{Kinetic model analysis}
\subsubsection{Kinetic model of xylopyranose decomposition}
\begin{figure}[htbp]
\centering
\includegraphics[width=0.9\columnwidth]{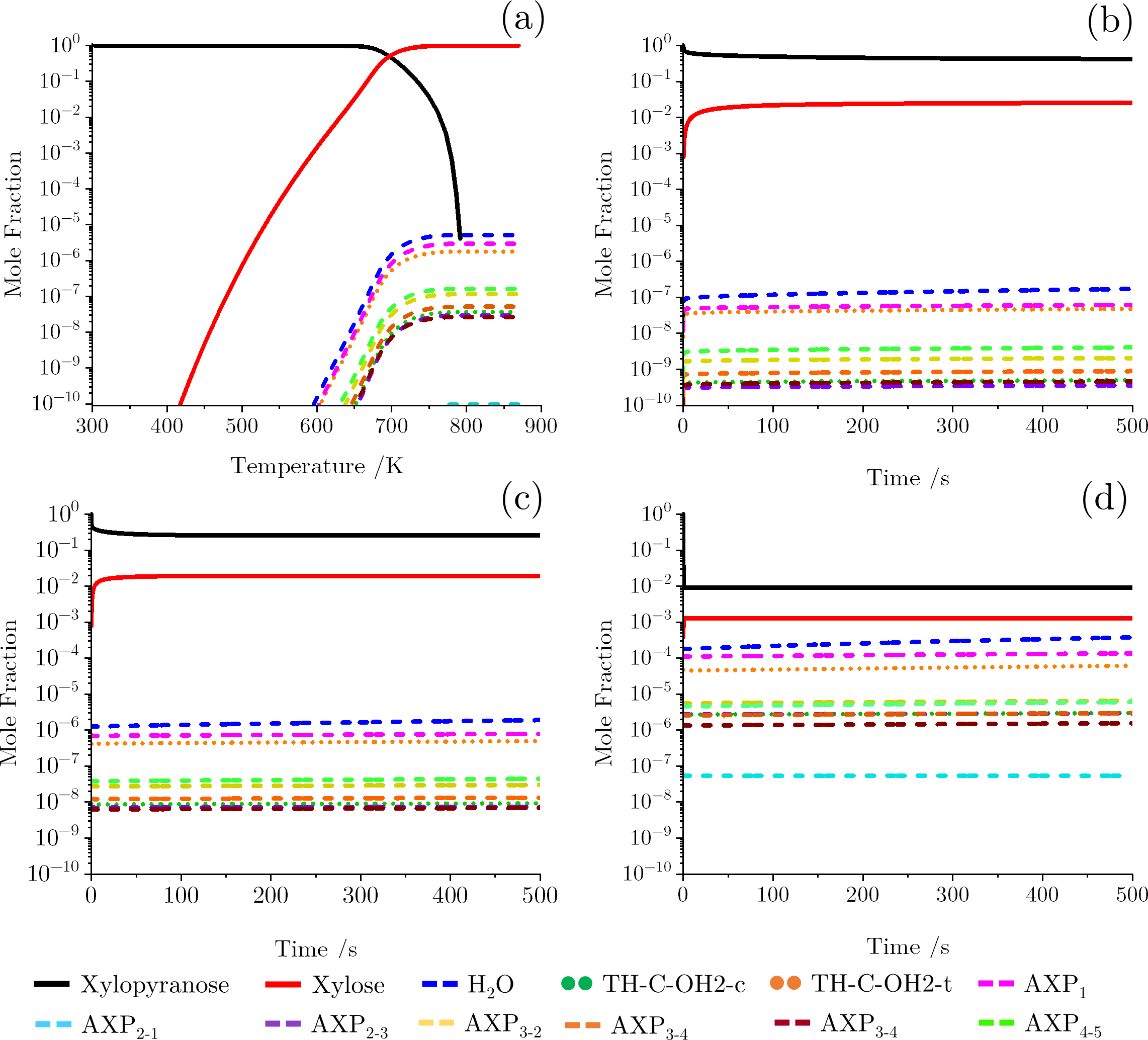} 
\caption{Mole fraction of chemical species produced from the initial competition decomposition reaction pathways of xylopyranose at 1 atm in: (a) TGA-like simulation with a heating rate of \SI{20}{\celsius\per\minute}. (b), (c), (d) Isobaric reactor at \SI{673.15}{\kelvin}, \SI{773.15}{\kelvin} and \SI{1073.15}{\kelvin} respectively as a function of time.} 
\label{fig:kinmodinitial}
\end{figure}

A chemical kinetic model was constructed for this system to allow for the competition between the initial decomposition step of \textbf{xylopyranose} to be studied using kinetic modelling and is reported in the Supporting Information materials. The Arrhenius pre-exponent ($A$) and activation energy ($E_a$) for each reaction pathway was obtained by fitting the temperature-dependent rate coefficients to the Arrhenius equation (Equation \ref{eq:arrhenius}) and are presented in Table \ref{tab:arrhpar}. Figure \ref{fig:kinmodinitial} (a) shows the results of a TGA-like simulation with a heating rate of \SI{20}{\celsius\per\minute} using this kinetic model. From this, one can observe that the thermal decomposition of \textbf{xylopyranose} begins at approximately \SI{670}{\kelvin}. This figure indicates that the ring opening reaction pathway leading to the \textbf{xylose} open-chain product is the preferred reaction pathway, as this product is present in large quantities, with the other products present in negligible concentrations. 
Figure \ref{fig:kinmodinitial} (b), (c) and (d) show the time evolution of \textbf{xylopyranose} in a fixed temperature isobaric reactor at \SI{673.15}{\kelvin}, \SI{773.15}{\kelvin} and \SI{1073.15}{\kelvin} respectively. These figures highlight that the ring opening pathway dominates at all temperatures, with the competition between pathways increasing as the temperature increases, as indicated in Figure \ref{fig:rates_xylo} (a). This is as one would expect by analysing the activation energies in Table \ref{tab:arrhpar} whereby the open chain reaction (\SI{181.8}{\kilo\joule\per\mol}) is approximately \SI{100}{\kilo\joule\per\mol} lower than the minimum activation energy observed among the other initial decomposition reaction channels. Based on this conclusion, it is reasonable to neglect the ring contraction and dehydration channels from further consideration. 

\subsubsection{Kinetic model of xylose decomposition}
\begin{figure}[htbp]
\centering
\includegraphics[width=0.9\columnwidth]{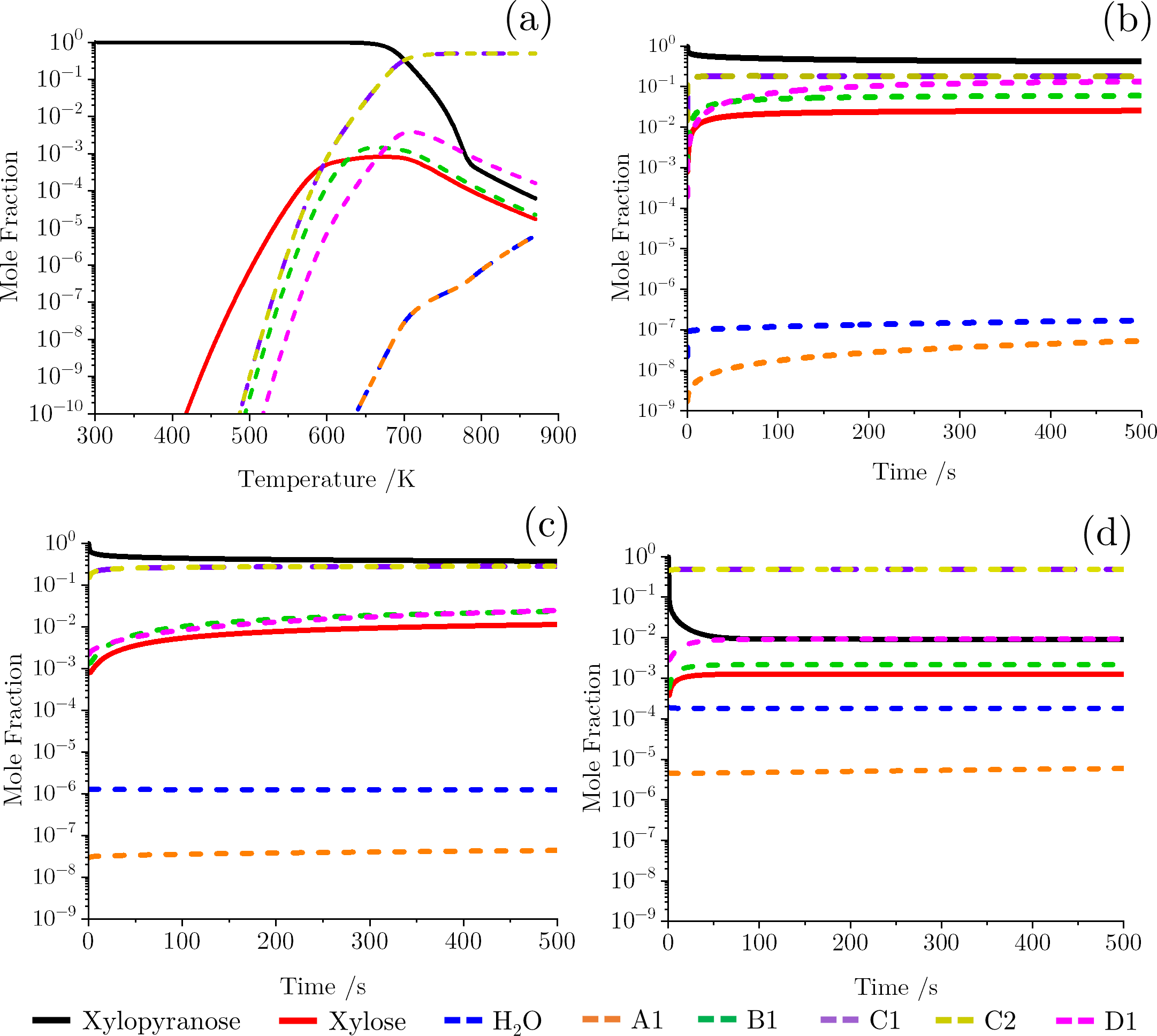} 
\caption{Mole fraction of chemical species produced from the second competition decomposition reaction pathways of xylopyranose at 1 atmosphere in: (a) TGA-like simulation with a heating rate of \SI{20}{\celsius\per\minute}. (b), (c), (d) Isobaric reactor at \SI{673.15}{\kelvin}, \SI{773.15}{\kelvin} and \SI{1073.15}{\kelvin} respectively as a function of time.} 
\label{fig:kinmodsecond}
\end{figure}

A second kinetic model is constructed by fitting the temperature-dependent reaction rate constants of Figure \ref{fig:rates_xylo} (b) to the Arrhenius equation. This kinetic model now describes the thermal degradation of \textbf{xylopyranose} to \textbf{xylose} and the subsequent competitive parallel reactions, namely dehydration (A), cyclization (B), C-C bond fission (C), and isomerization (D), as depicted in Figure \ref{fig:genscheme}. The results of the TGA-like simulation, shown in Figure \ref{fig:kinmodsecond} (a) shows that \textbf{xylopyranose} decomposes to \textbf{xylose}, as in Figure \ref{fig:kinmodinitial} (a), however the \textbf{xylose} now breaks down through these new reaction pathways. At low temperatures, \SIrange[range-phrase=--,range-units=single]{400}{600}{\kelvin}, \textbf{xylose} decomposes primarily to \textbf{C1} and \textbf{C2}, however a sizeable quantity of \textbf{B1} and \textbf{D1} is also produced. As the temperature increases however, the concentration of \textbf{B1} and \textbf{D1} decreases. At higher temperatures, \SIrange[range-phrase=--,range-units=single]{800}{900}{\kelvin}, the concentration of \textbf{A1} and \textbf{\ce{H2O}} begins to increase as the dehydration of xylose pathway begins to compete with that of the other pathways. This is as expected from analysis of the reaction rate constants only, e.g. Figure \ref{fig:rates_xylo} (b). Additionally, it is interesting to note that the concentration of \textbf{xylopyranose} remains orders of magnitude higher at high temperatures compared to in Figure \ref{fig:kinmodinitial} (a). This is likely due to the removal of the \textbf{xylopyranose} ring contraction and dehydration pathways which become competitive at high temperatures, as shown in Figure \ref{fig:rates_xylo} (a). 
Figure \ref{fig:kinmodsecond} (b), (c) and (d) show the time evolution of \textbf{xylopyranose} in a fixed temperature isobaric reactor at \SI{673.15}{\kelvin}, \SI{773.15}{\kelvin} and \SI{1073.15}{\kelvin} respectively. The results of these simulations further reinforce the conclusions drawn from Figure \ref{fig:kinmodsecond} (a) that the C-C bond fission reaction pathway dominates at all temperatures, producing \textbf{C1} and \textbf{C2}, with all other products having mole fractions orders of magnitudes lower.
This is a demonstration of the method of iteratively evaluating the competition between parallel reaction pathways, using temperature-dependent reaction rate constants and kinetic modelling, and retaining only those that contribute appreciably to the formation of the reaction products.  In our view, this approach has the potential to establish a general strategy for systematically and iteratively exploring reactive pathways in pyrolytic systems, opening up the possibility of automated reaction mechanism construction.

\subsubsection{Reaction flux analysis}
\begin{figure}[htbp]
\centering
\includegraphics[width=0.8\columnwidth]{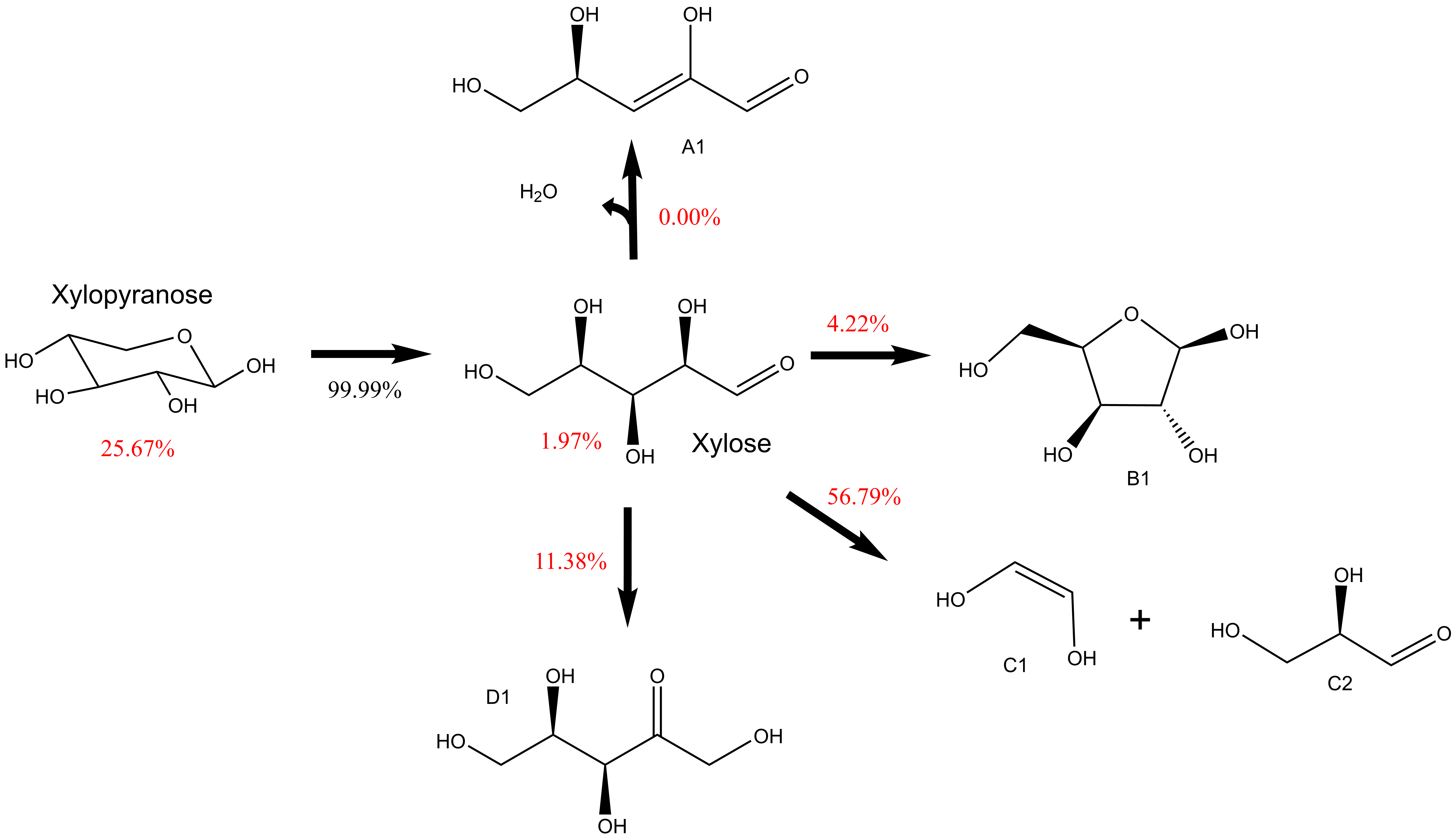} 
\caption{Flux diagram depicting $\beta$-D-xylopyranose decomposition at \SI{773.15}{\kelvin}. Red numbers represent branching ratios calculated with the kinetic model, including the second set of competitive pathways, while black number indicates the first competition leading to D-xylose open chain formation.} 
\label{fig:flux}
\end{figure}

Figure \ref{fig:flux} illustrates the chemical flux analysis of xylopyranose at \SI{773.15}{\kelvin} and \SI{1}{\atmosphere}, representing a critical set of test conditions explored in our study. The flux analysis offers valuable insights into the pyrolysis of xylopyranose under fast pyrolysis conditions. Our analysis reveals that nearly 99\% of xylopyranose undergoes decomposition into the xylose open chain, with only trace amounts of xylopyranose participating in water-loss reactions.

When we incorporate the second competitive reaction pathways into the model (highlighted by the red numbers), a distinct shift in product distribution is observed. Approximately 57\% of the xylopyranose is transformed into \ce{C{1} + C{2}}, followed by about 11\% forming xylulose (D1), and 4\% yielding xylofuranose (B1). It is worth noting that a substantial portion of xylose (circa 26\%) goes back to xylopyranose, due to its thermodynamic stability.

\section{Conclusions}
In summary, the gas-phase pyrolytic reactivity of $\beta$-D-xylopyranose, which represents a fundamental structural motif of hemicellulose, has been investigated. A combined quantum chemistry/chemical kinetics analysis has been performed on literature selected reaction pathways, employing a higher level of theory. Electronic energies and finite temperature corrections were recomputed at a higher level of theory, namely DLPNO-CCSD(T)-F12/cc-pVTZ-F12//M06-2X/6-311++G(d,p). Rate coefficients for specific reaction channels have been computed by means of conventional TST.
The interplay between thermodynamics and kinetics leads to the identification of xylopyranose ring opening pathway as the main initial reactive channel. Accordingly, our electronic structure calculations concentrate exclusively on the reactive pathways associated with the ring opening reaction.
Our calculations are in good agreement with previously reported works. From the open chain xylose, a series of thermodinamically stable, key products can be formed: furfural, anhydro-D-xylopyranose, glycolaldehyde, dihydrofuran-3(2H)-one. In particular furfural revests a key role among renewable chemicals, since it can be transformed to ethyl levulinate (EL), which is a an important biofuel additive.\cite{furfural,furfuralEL}
For the first time, enthalpies of formation, heat capacities at constant pressure, and reaction barriers are reported for the key species involved in the xylose pyrolytic system.
By applying Transition State Theory and solving the kinetic model through Cantera simulations, different pyrolytic reactive regimes of xylopyranose have been replicated, namely fixed temperature and increasing temperature ones (TGA-like). While the kinetic model remains incomplete and will undergo future expansion, from a methodological perspective, the proposed approach can pave the way to an iterative exploration of potential energy surfaces for reactive systems of this nature. Indeed the idea is to include new reactions into the model only if their contribution to the total reaction flux is above a certain threshold, otherwise, these reactions are excluded. 

In perspective, this iterative interplay between PES exploration and kinetic model validation could be automatized through the use of codes that automatically search for reaction mechanisms.\cite{automekin,automekin2015a,automekin2015b,chemtrayzer2015,chemtrayzer2017,chemtrayzer2018}
Future enhancements to the model will involve the incorporation of additional reaction pathways and a comparison of the kinetic model's results with experimental data.

\begin{acknowledgement}
This publication has emanated from research conducted with the financial support of the Ryanair Sustainable Aviation Research Center at Trinity College Dublin, the European Union through the European Research Council, Mod-L-T, action number 101002649, and the Science Foundation Ireland (SFI) under grant number 12/RC/2278\_2. This publication is also co-funded under the European Regional Development Fund under the AMBER award. The authors wish to acknowledge the Irish Centre for High-End Computing (ICHEC) for the provision of computational facilities and support. JL thanks Dr. Bernardo Ballotta for engaging in invaluable discussions that have significantly improved the quality of the draft version of the paper.
\end{acknowledgement}

\begin{suppinfo}
Cantera kinetic model input file; geometries of all the stationary points optimized at M06-2X/6-311++G(d,p).
\end{suppinfo}

\bibliography{biblio}

\end{document}


\section{Cantera kinetic model}
\begin{tiny}
\begin{verbatim}
# Kinetic model of the pyrolysis of xylopyranose
units(time='s', quantity='mol', act_energy='cal/mol')

ideal_gas(name='biomass',
          elements="C H O N Ar He",
          species="""XYLPYRANOSE XYLOSE THCOH2-c H2O THCOH2-t
                     AXP1 AXP2-1 AXP2-3 AXP3-2 AXP3-4 AXP4-3 AXP4-5
					 A1 B1 C1 C2 D1
                     """,
          reactions='all',
          initial_state=state(temperature=300.0, pressure=OneAtm))


#-----------------------------------------------------------------------------------------------------------------------
# Species data
#-----------------------------------------------------------------------------------------------------------------------
#---------XYLOPYRANOSE SPECIES-------------------------------------------------------------------------------------------------
species(name='XYLPYRANOSE',
        atoms='C:5 H:10 O:5',
        thermo=(NASA([300.00, 1406.00],
                     [ 2.49780875E+01, 2.69185990E-02, -9.05065171E-06, 1.38943725E-09, -7.99958269E-14, -1.17895625E+05, -1.08736372E+02 ]),
                NASA([1406.00, 5000.00],
                     [-2.65400291E+00, 9.69870855E-02, -7.73842184E-05, 3.17132644E-08, -5.21911434E-12, -1.08970489E+05, 3.75072158E+01 ])))

species(name='XYLOSE',
        atoms='C:5 H:10 O:5',
        thermo=(NASA([300.00, 1406.00],
                     [2.49986873E+01, 2.69495939E-02, -9.06958437E-06, 1.39305827E-09, -8.02276646E-14, -1.15594775E+05, -1.07504846E+02 ]),
                NASA([1406.00, 5000.00],
                     [-2.23787169E+00, 9.72741171E-02, -7.93710822E-05, 3.35153576E-08, -5.69606739E-12, -1.06888770E+05, 3.62563810E+01 ])))

species(name='AXP1',
        atoms='C:5 H:8 O:4',
        thermo=(NASA([300.00, 1406.00],
                     [2.18031893E+01, 2.21890022E-02, -7.46463135E-06, 1.14638839E-09, -6.60199283E-14, -7.98351713E+04, -9.13650716E+01 ]),
                NASA([1406.00, 5000.00],
                     [-2.49586345E+00, 8.55649305E-02, -7.13237938E-05, 3.04787418E-08, -5.20842323E-12, -7.21544244E+04, 3.66279212E+01 ])))

species(name='AXP2-1',
        atoms='C:5 H:8 O:4',
        thermo=(NASA([300.00, 1406.00],
                     [2.18031893E+01, 2.21890022E-02, -7.46463135E-06, 1.14638839E-09, -6.60199283E-14, -7.98351713E+04, -9.13650716E+01 ]),
                NASA([1406.00, 5000.00],
                     [-2.49586345E+00, 8.55649305E-02, -7.13237938E-05, 3.04787418E-08, -5.20842323E-12, -7.21544244E+04, 3.66279212E+01 ])))
                    
species(name='AXP2-3',
        atoms='C:5 H:8 O:4',
        thermo=(NASA([300.00, 1406.00],
                     [2.16444065E+01, 2.23573199E-02, -7.53005766E-06, 1.15736608E-09, -6.66900432E-14,  -8.34832888E+04,  -9.08417686E+01 ]),
                NASA([1406.00, 5000.00],
                     [-3.24423062E+00, 8.75755299E-02, -7.36280218E-05, 3.17123452E-08, -5.45800091E-12,  -7.56412771E+04,   4.01578934E+01 ])))
                      
species(name='AXP3-2',
        atoms='C:5 H:8 O:4',
        thermo=(NASA([300.00, 1406.00],
                     [2.16854517E+01, 2.23146293E-02, -7.51367629E-06, 1.15464242E-09, -6.65248819E-14,  -8.31608253E+04,  -9.10625813E+01 ]),
                NASA([1406.00, 5000.00],
                     [-3.21529048E+00, 8.76696612E-02, -7.38678163E-05, 3.18823852E-08, -5.49725094E-12,  -7.53251621E+04,   3.99641068E+01 ])))

species(name='AXP3-4',
        atoms='C:5 H:8 O:4',
        thermo=(NASA([300.00, 1406.00],
                     [2.17439640E+01, 2.22495462E-02, -7.48790311E-06, 1.15027930E-09, -6.62572466E-14, -8.37159932E+04, -9.14060963E+01 ]),
                NASA([1406.00, 5000.00],
                     [-3.04797056E+00, 8.73013918E-02, -7.34929863E-05, 3.16887437E-08, -5.45774968E-12, -7.59152261E+04,  3.90501979E+01 ])))
                    
species(name='AXP4-3',
        atoms='C:5 H:8 O:4',
        thermo=(NASA([300.00, 1405.00],
                     [2.16699636E+01, 2.23858974E-02, -7.55080493E-06, 1.16172105E-09, -6.69882638E-14,  -8.33168771E+04,  -9.09822391E+01 ]),
                NASA([1405.00, 5000.00],
                     [-2.55688787E+00, 8.48849273E-02, -6.98233180E-05, 2.94671545E-08, -4.98436440E-12,  -7.55849082E+04,   3.68836223E+01 ])))
                     
species(name='AXP4-5',
        atoms='C:5 H:8 O:4',
        thermo=(NASA([300.00, 1408.00],
                     [2.17488415E+01, 2.22158424E-02, -7.46983525E-06, 1.14679591E-09, -6.60278306E-14,  -8.36944064E+04,  -9.21519737E+01 ]),
                NASA([1408.00, 5000.00],
                     [-3.94902499E+00, 9.05978175E-02, -7.78414305E-05, 3.41306697E-08, -5.95450857E-12,  -7.57089384E+04,   4.27246754E+01 ])))
                     
species(name='THCOH2-c',
        atoms='C:5 H:8 O:4',
        thermo=(NASA([300.00, 1403.00],
                     [2.14595236E+01,  2.28324837E-02, -7.76248131E-06,  1.20064221E-09, -6.94887961E-14,  -8.16346773E+04,  -8.92399779E+01 ]),
                NASA([1403.00, 5000.00],
                     [-2.78418309E+00,  8.34511371E-02, -6.60365291E-05,  2.67147031E-08, -4.34315075E-12,  -7.37051845E+04,   3.93969674E+01 ])))
                     
species(name='THCOH2-t',
        atoms='C:5 H:8 O:4',
        thermo=(NASA([300.00, 1404.00],
                     [2.15833423E+01,  2.26832909E-02, -7.70147009E-06, 1.19013876E-09, -6.88379839E-14,  -8.13232283E+04, -8.97092399E+01  ]),
                NASA([1404.00, 5000.00],
                     [-2.63012046E+00,  8.35816307E-02, -6.66370208E-05, 2.71752828E-08, -4.45136295E-12,  -7.34404439E+04,  3.86389919E+01   ])))
                     
species(name='H2O',
        atoms='C:0 H:2 O:1',
        thermo=(NASA([300.00, 1612.00],
                     [2.69405642E+00, 2.91342241E-03, -8.44269107E-07,  1.15325994E-10, -6.05732068E-15,  -3.00545010E+04, 6.81403842E+00  ]),
                NASA([1612.00, 5000.00],
                     [3.88299529E+00, 6.19091267E-05,  1.72704490E-06, -9.26315808E-10,  1.55265050E-13,  -3.04492571E+04, 4.71351995E-01   ])))
					 
species(name='A1',
        atoms='C:5 H:8 O:4',
        thermo=(NASA([300.00, 1403.00],
                     [2.20765007E+01, 2.19812856E-02, -7.39644389E-06,  1.13595359E-09, -6.54160504E-14,  -7.75356836E+04, -8.83850414E+01  ]),
                NASA([1403.00, 5000.00],
                     [8.29963867E-01, 7.60319897E-02, -6.05917153E-05,  2.50815626E-08, -4.19657181E-12,  -7.06579710E+04,  2.40553993E+01   ])))


species(name='B1',
        atoms='C:5 H:10 O:5',
        thermo=(NASA([300.000, 1408.000],
                     [2.50927027E+01, 2.67628658E-02, -8.98388513E-06, 1.37764625E-09, -7.92535393E-14, -1.16531690E+05, -1.08786494E+02]),
                NASA([1408.000, 5000.000],
                     [-3.59647221E+00, 1.01283289E-01, -8.36404930E-05, 3.54322094E-08, -6.00480758E-12, -1.07443120E+05, 4.24216425E+01])))

species(name='C1',
        atoms='C:2 H:4 O:2',
        thermo=(NASA([300.000, 1411.000],
                     [9.66817211E+00, 9.75467333E-03, -3.21136261E-06, 4.85719100E-10, -2.76660724E-14, -3.71530983E+04, -2.56112323E+01]),
                NASA([1411.000, 5000.000],
                     [5.27893963E-01, 3.47936408E-02, -2.98233333E-05, 1.33600089E-08, -2.39248650E-12, -3.43745531E+04, 2.21236995E+01])))

species(name='C2',
        atoms='C:3 H:6 O:3',
        thermo=(NASA([300.000, 1400.000],
                     [1.45765120E+01, 1.58996837E-02, -5.35877165E-06, 8.23867005E-10, -4.74771981E-14, -6.30477022E+04, -4.97688546E+01]),
                NASA([1400.000, 5000.000],
                     [1.64191594E+00, 4.62318980E-02, -3.23479381E-05, 1.16531685E-08, -1.69977078E-12, -5.86019217E+04, 1.95982465E+01])))

species(name='D1',
        atoms='C:5 H:10 O:5',
        thermo=(NASA([300.000, 1405.000],
                     [2.52982330E+01, 2.66040581E-02, -8.93068163E-06, 1.36935573E-09, -7.87666925E-14, -1.14902239E+05, -1.06070638E+02]),
                NASA([1405.000, 5000.000],
                     [1.69037888E-01, 8.99751114E-02, -7.05400088E-05, 2.86955402E-08, -4.71800379E-12, -1.06726874E+05, 2.70898135E+01])))


#-----------------------------------------------------------------------------------------------------------------------
# Reaction data
#-----------------------------------------------------------------------------------------------------------------------
#----------XYLOPYRANOSE REACTIONS------------------------------------------------------------------------------------------          


# Reaction || 1
# Ring opening
reaction('XYLPYRANOSE <=> XYLOSE', [2.53e12, 0.0, 43451.2908])

# Reaction || 2
# Ring contraction to (THCOH2-c) dihydroxytetrahydrofuran-2-carbaldehyde isomer cis.
#reaction('XYLPYRANOSE <=> THCOH2-c + H2O', [7.08e13, 0.0, 72920.7306])

# Reaction || 3
# Ring contraction to (THCOH2-t) dihydroxytetrahydrofuran-2-carbaldehyde isomer cis.
#reaction('XYLPYRANOSE <=> THCOH2-t + H2O', [6.42e13, 0.0, 67112.8848])

# Reaction || 4
# Dehydration elimination at 1 position to 1-anhydro xylopyranose, AXP1. 
#SD - change to D1
#reaction('XYLPYRANOSE <=> AXP1 + H2O', [4.6e14, 0.0, 69240.0382 ])

# Reaction || 5
# Dehydration elimination at 2 position to 2-anhydro xylopyranose-1, AXP2-1.
# D2-1, meaning loss of OH at 2 and loss of H at 1.
#reaction('XYLPYRANOSE <=> AXP2-1 + H2O', [2.76e14, 0.0, 83628.1994])

# Reaction || 6
# Dehydration elimination at 2 position to 2-anhydro xylopyranose-3, AXP2-3.
#reaction('XYLPYRANOSE <=> AXP2-3 + H2O', [1.23e14, 0.0, 74044.0588])

# Reaction || 7
# Dehydration elimination at 3 position to 3-anhydro xylopyranose-2, AXP3-2.
#reaction('XYLPYRANOSE <=> AXP3-2 + H2O', [4.58e13, 0.0, 70602.3724])

# Reaction || 8
# Dehydration elimination at 3 position to 3-anhydro xylopyranose-4, AXP3-4.
#reaction('XYLPYRANOSE <=> AXP3-4 + H2O', [2.32e13, 0.0, 70793.5772])

# Reaction || 9
# Dehydration elimination at 4 position to 4-anhydro xylopyranose-3, AXP4-3.
#reaction('XYLPYRANOSE <=> AXP4-3 + H2O', [1.26e13, 0.0, 70865.2790])

# Reaction || 10
# Dehydration elimination at 4 position to 4-anhydro xylopyranose-5, AXP4-5.
#reaction('XYLPYRANOSE <=> AXP4-5 + H2O', [8.83e12, 0.0, 67710.3998])

# Reaction || 11
# Dehydration elimination from XYLOSE open chain, at 3 position 	
reaction('XYLOSE <=> A1 + H2O', [1.17e15, 0.0, 65344.2404])

# Reaction || 12
# Cyclization reaction from XYLOSE open chain to five member intermediate B1 (xylofuranose)
reaction('XYLOSE <=> B1', [4.17e12, 0.0, 37380.5384])

# Reaction || 13
# Carbon-carbon bond fission reaction from XYLOSE open chain to C1 (enol isomer of glycolaldehyde) and C2 (glyceraldehyde)
reaction('XYLOSE <=> C1 + C2', [1.76e14, 0.0, 39794.4990])

# Reaction || 14
# Isomerization reaction from XYLOSE open chain to D1 (D-xylulose)
reaction('XYLOSE <=> D1', [4.38e13, 0.0, 43666.3962])
\end{verbatim}
\end{tiny}
\section{Geometries of stationary points}
\subsection{Xylopyranose decompositions}

\subsubsection{Ring opening}
\paragraph{Xylopyranose}
\begin{verbatim}

0 1
 C                 -1.38315700   -0.55311700   -0.29076300
 C                 -0.55952400   -1.72670300    0.23383900
 C                  1.44159700   -0.46690700    0.25020300
 C                  0.72625100    0.79532500   -0.21250600
 C                 -0.73908100    0.73151100    0.19091500
 H                 -0.59924600   -1.73198400    1.33267900
 H                 -0.94607800   -2.67940500   -0.13073300
 H                 -1.38574100   -0.56202900   -1.38734200
 H                  1.45715600   -0.49706600    1.35260400
 H                  0.80100300    0.83544200   -1.30756000
 H                 -0.79856600    0.74964500    1.28998200
 O                  0.76988300   -1.62118800   -0.23075300
 O                  2.71618000   -0.46676300   -0.28026000
 H                  3.16424200   -1.26710100    0.00864200
 O                  1.35010000    1.90569800    0.38384200
 H                  0.81527900    2.67796400    0.17189000
 O                 -1.36555400    1.87940900   -0.34383800
 H                 -2.29659200    1.84373300   -0.10187300
 O                 -2.70700100   -0.55343100    0.22025200
 H                 -3.23683600   -1.19965000   -0.25235700
\end{verbatim}
\paragraph{Xylose}
\begin{verbatim}
0 1
 C                 -0.84707700   -0.92723300   -0.01342800
 C                 -2.31335000   -0.52233800    0.12679500
 C                  1.82690300    0.63350200   -0.62247100
 C                  1.52996900   -0.31227600    0.53247600
 C                  0.05984000   -0.21522600    0.99974100
 H                 -2.93694600   -1.21619500   -0.43833800
 H                 -2.61093000   -0.56076800    1.18008400
 H                 -0.75969100   -2.00429100    0.15615300
 H                  2.31714800    0.16858500   -1.49751400
 H                  2.17747500   -0.00770100    1.36064700
 H                 -0.02491200   -0.74318000    1.95476500
 O                 -2.53128200    0.75744200   -0.43292200
 O                  1.58714900    1.81456700   -0.58966500
 H                 -2.08409900    1.38511000    0.14834600
 O                  1.88136100   -1.63096100    0.19937600
 H                  1.42708100   -1.86093100   -0.62064500
 O                 -0.33601000    1.12714500    1.23456100
 H                  0.23687400    1.72321500    0.72888300
 O                 -0.39612000   -0.67822300   -1.33953800
 H                 -0.92048900    0.05782500   -1.68555400

\end{verbatim}
\paragraph{TS-Xylose}
\begin{verbatim}
0 1
 O                 -2.55104400    0.02081100   -0.11237300
 O                  0.42262000    2.04978100   -0.09199000
 O                 -0.50633500   -1.00624400    1.62008400
 O                  2.36281200    0.17529300   -0.28462900
 O                  0.76857800   -0.96914200   -1.12418200
 C                 -0.39269400    1.03930800    0.44548100
 C                  0.42638700   -0.02835300    1.17464200
 C                  1.46894700   -0.64698300    0.24869300
 C                 -0.65807000   -0.83318100   -1.31975600
 C                 -1.29264900    0.39569400   -0.63305200
 H                 -2.38785600   -0.54789600    0.64969400
 H                  1.32375000    1.70346900   -0.20731300
 H                 -0.09474200   -1.59325900    2.25922800
 H                  1.55763600   -0.19746700   -1.35350800
 H                 -1.07699600    1.49803300    1.16526900
 H                  0.96779900    0.42853300    2.01015200
 H                  1.82440500   -1.62662900    0.59801000
 H                 -1.49074500    1.17743900   -1.36656400
 H                 -0.80617500   -0.82648200   -2.39833500
 H                 -1.10165400   -1.73863700   -0.90795700

\end{verbatim}
\subsubsection{Dehydrations}
\paragraph{\ce{AXP_{1}}}
\begin{verbatim}
0 1
 C                 -0.42769500   -0.93194800    0.21140800
 C                  1.09433000   -0.82120600    0.09253000
 C                 -1.03695900    0.43646000    0.19650700
 C                  1.42761600    0.25649200   -0.92620300
 O                  1.68503400   -0.53657100    1.34098800
 H                  1.47809900   -1.78650400   -0.24096400
 O                 -0.88730400   -1.72205200   -0.87413400
 H                 -0.66542200   -1.41254900    1.16765500
 C                 -0.33896300    1.53700000   -0.08441200
 O                 -2.37689500    0.41070000    0.49218600
 O                  0.98218300    1.52307900   -0.44418900
 H                  2.50263300    0.35222200   -1.06774300
 H                  0.92969800    0.03175400   -1.87533200
 H                  1.58342700    0.40587100    1.51439300
 H                 -1.84801900   -1.67536800   -0.89229000
 H                 -0.75047000    2.53914000   -0.04032200
 H                 -2.76407000    1.28338800    0.37681900

\end{verbatim}
\paragraph{\ce{AXP_{2-1}}}
\begin{verbatim}
0 1
 C                 -1.32171300    0.24653900   -0.07201500
 C                 -0.40459100    1.05997800   -0.61483100
 O                 -2.64099800    0.40267200   -0.25917900
 C                  1.05734300    0.79829400   -0.41362300
 H                 -0.75191600    1.88941100   -1.21536100
 O                 -1.06309200   -0.84130900    0.67545700
 H                 -3.09844500   -0.26453400    0.26509000
 C                  1.26352800   -0.66396100   -0.01659300
 O                  1.63668600    1.58027700    0.63254300
 H                  1.60429100    0.98150100   -1.34406800
 C                  0.29864300   -1.00117900    1.10779300
 O                  1.11064500   -1.53217200   -1.11646800
 H                  2.28374700   -0.78775600    0.34986000
 H                  1.12854200    2.39035400    0.71830400
 H                  0.38468300   -2.04380400    1.40553900
 H                  0.47651300   -0.33850300    1.95840000
 H                  0.26739600   -1.34043100   -1.54097900

\end{verbatim}
\paragraph{\ce{AXP_{2-3}}}
\begin{verbatim}
0 1
 C                  0.82106800    0.81466400    0.09098500
 C                  1.06047300   -0.62562000    0.46243800
 C                 -0.36628900    1.25205300   -0.31609500
 C                 -0.26380800   -1.36181800    0.57971500
 O                  1.96559000   -1.21854800   -0.45991400
 H                  1.57062400   -0.66151000    1.43006800
 O                  1.89672400    1.63969100    0.18425700
 C                 -1.51168700    0.30153900   -0.51106800
 H                 -0.52419100    2.29622400   -0.55447800
 O                 -1.10015600   -1.04695800   -0.52135100
 H                 -0.09648000   -2.43818800    0.55321000
 H                 -0.75845600   -1.08825000    1.51719800
 H                  1.52680000   -1.24671900   -1.31778500
 H                  2.69680100    1.10259800    0.17398200
 O                 -2.46777200    0.53605500    0.49398100
 H                 -1.95770800    0.45204600   -1.49895500
 H                 -3.25101100    0.01697300    0.28513400

\end{verbatim}
\paragraph{\ce{AXP_{3-2}}}
\begin{verbatim}
0 1
 C                  0.74431000    0.84586400    0.02796400
 C                  0.99994100   -0.56930700    0.48525800
 C                 -0.45285500    1.26198500   -0.37048900
 O                  1.99108400   -1.08412100   -0.38032200
 O                 -0.14577200   -1.35526900    0.53053700
 H                  1.37976400   -0.54900400    1.51342700
 O                  1.82249000    1.66796100    0.11692700
 C                 -1.59247700    0.28418000   -0.48790800
 H                 -0.60915100    2.30368700   -0.62405500
 H                  2.27089800   -1.94212500   -0.04651000
 C                 -1.04059900   -1.13151600   -0.55937100
 H                  2.61711900    1.13759600   -0.01526800
 O                 -2.54534200    0.42776500    0.55423400
 H                 -2.14566000    0.48266600   -1.40925900
 H                 -1.84204400   -1.85890800   -0.44226300
 H                 -0.52182000   -1.29568800   -1.50932600
 H                 -2.07870400    0.32385100    1.38953100
\end{verbatim}
\paragraph{\ce{AXP_{3-4}}}
\begin{verbatim}
0 1
 C                  0.97667800    0.87959300   -0.38295000
 C                  1.28919200   -0.56413500    0.00727000
 C                 -0.50192100    1.07375200   -0.55894000
 O                  0.43803300   -0.99670800    1.04600600
 O                  1.16589500   -1.36547600   -1.13360300
 H                  2.29358900   -0.60964900    0.43408200
 O                  1.51921800    1.76542400    0.58454500
 H                  1.50838800    1.09684500   -1.31073100
 C                 -1.36023200    0.18173500   -0.06972900
 H                 -0.84749700    1.96733700   -1.06958100
 C                 -0.92572700   -1.04714200    0.67146000
 H                  1.49354100   -2.24532600   -0.92619900
 H                  1.00123100    1.66265100    1.38961700
 O                 -2.71253600    0.23016200   -0.17696900
 H                 -1.50263500   -1.13587200    1.59310100
 H                 -1.13132900   -1.92624600    0.05016900
 H                 -2.96810100    0.98024000   -0.72295100

\end{verbatim}
\paragraph{\ce{AXP_{4-3}}}
\begin{verbatim}
0 1
 C                  0.15470200   -0.77477700   -0.31725100
 C                 -1.13666600   -0.22645400    0.27745100
 C                  1.25868100    0.22433900   -0.09144400
 O                 -1.38015100    1.03903500   -0.26928200
 O                 -2.21815100   -1.05569600   -0.01117900
 H                 -1.04546800   -0.18575600    1.37058900
 O                  0.54278000   -1.98531400    0.30475700
 H                  0.00420700   -0.93122700   -1.39556800
 C                  0.98520700    1.50862400    0.11331400
 O                  2.52181300   -0.26564400   -0.13719400
 C                 -0.42967000    2.01670800    0.13857900
 H                 -2.52020900   -0.83508400   -0.89988500
 H                 -0.17692400   -2.61753800    0.21075800
 H                  1.78887200    2.21516600    0.28326700
 H                  2.47304600   -1.21879500    0.00963700
 H                 -0.55863000    2.85581700   -0.54692900
 H                 -0.68874500    2.36772600    1.14743300

\end{verbatim}
\paragraph{\ce{AXP_{4-5}}}
\begin{verbatim}
0 1
 C                 -0.58674700   -0.04496300    1.50676100
 C                  0.72310600   -0.26833200    1.58337500
 C                  0.99278500   -0.21668000   -0.75847100
 C                 -0.19198500    0.74448200   -0.80940400
 C                 -1.25941800    0.31137700    0.20630600
 H                  1.24334700   -0.50214500    2.50345700
 H                 -1.18351100   -0.13011600    2.40559900
 H                  1.78958500    0.09974200   -1.43371600
 H                 -0.62270300    0.72803700   -1.81085500
 H                 -1.91878500    1.16929900    0.35580800
 O                  1.59260700   -0.21203400    0.52686900
 O                  0.51413900   -1.49555600   -1.07333900
 H                  1.24735500   -2.11786600   -1.05240000
 O                  0.26026500    2.05730300   -0.56885600
 H                  0.64021800    2.09428600    0.31596300
 O                 -2.08541900   -0.72663200   -0.28306300
 H                 -1.51467800   -1.48120100   -0.46813900

\end{verbatim}
\paragraph{\ce{H2O}}
\begin{verbatim}
0 1
 O                  0.00000000    0.00000000    0.11661200
 H                  0.00000000    0.76156400   -0.46644700
 H                  0.00000000   -0.76156400   -0.46644700

\end{verbatim}
\paragraph{\ce{TS-AXP_{1}}}
\begin{verbatim}
0 1
 C                  0.60082500    0.85490700   -0.31073700
 C                  1.30722300   -0.31457800    0.35029200
 C                 -0.87736200    0.57298200   -0.22960800
 C                  0.96573800   -1.60610200   -0.37560500
 O                  2.70803300   -0.19768700    0.27595900
 H                  0.96905200   -0.36421900    1.39247800
 O                  0.96152300    2.01372600    0.41050600
 H                  0.94000900    0.94913500   -1.35322100
 C                 -1.25822800   -0.76453000   -0.60006300
 O                 -1.67146100    1.59613200   -0.78349900
 O                 -0.44661000   -1.76879900   -0.65339600
 H                  1.25079900   -2.47634800    0.21075400
 H                  1.48038600   -1.63260100   -1.33775400
 H                  2.94019000    0.66582700    0.63483800
 H                  0.42148100    2.74278500    0.08926300
 H                 -2.25610600   -0.98653200   -0.95628100
 H                 -2.38401900    1.78236800   -0.16475700
 H                 -1.24998100    0.31234200    0.93448500
 O                 -2.06542100   -0.68555100    1.41827500
 H                 -2.42950500   -1.11140000    2.20176100
\end{verbatim}
\paragraph{\ce{TS-AXP_{2-1}}}
\begin{verbatim}
0 1
 C                 -0.81665700    0.81866400   -0.00303700
 C                 -1.38136600   -0.58508800   -0.18705000
 C                  0.68312800    0.82234600   -0.24380500
 C                 -0.49799800   -1.58077600    0.54986500
 O                 -2.68911100   -0.66859300    0.32505600
 H                 -1.35901500   -0.84630700   -1.25717500
 O                 -1.49724500    1.75675500   -0.81695000
 H                 -1.01565500    1.12635400    1.02469600
 C                  1.43045600   -0.39222900   -0.08227900
 H                  1.05372000    1.59297800   -0.91551400
 O                  0.76487600   -1.63066400   -0.07140400
 H                 -0.92819800   -2.57929500    0.48962100
 H                 -0.41066200   -1.29144900    1.60688000
 H                 -3.18513200    0.09096700    0.00223000
 H                 -1.35719300    1.52636600   -1.74126800
 O                  2.55478700   -0.47351000   -0.89569700
 H                  3.19403400   -1.04464000   -0.46263600
 H                  1.70140100    0.42677300    1.15350000
 O                  1.33947700    1.54957900    1.29766400
 H                  2.01904900    2.23221400    1.18815000

\end{verbatim}
\paragraph{\ce{TS-AXP_{2-3}}}
\begin{verbatim}
0 1
 C                 -0.83666600    0.55052500   -0.31685300
 C                 -1.14708500   -0.92565800   -0.30671600
 C                  0.50731100    1.02222800   -0.24986500
 C                  0.11546400   -1.75596600   -0.10416600
 O                 -2.17109200   -1.24212400    0.63498300
 H                 -1.58031500   -1.19402200   -1.27446700
 O                 -1.66293000    1.33770600   -1.15252700
 C                  1.58429500    0.05412900    0.21026300
 H                  0.85183200    1.80814600   -0.91600400
 O                  1.04478900   -1.10846000    0.76590300
 H                 -0.13305500   -2.70437600    0.36952300
 H                  0.59434700   -1.94389100   -1.06929500
 H                 -1.88064100   -0.91788000    1.49360000
 H                 -2.55734700    1.29155300   -0.80518500
 O                  2.35564700   -0.20947800   -0.93423900
 H                  2.18216200    0.50262700    1.00366000
 H                  3.08606900   -0.77924000   -0.67373800
 H                 -0.67425800    1.28646200    0.95407700
 O                  0.26628500    1.98011900    1.24399800
 H                  0.10970000    2.91696700    1.05691200

\end{verbatim}
\paragraph{\ce{TS-AXP_{3-2}}}
\begin{verbatim}
0 1
 C                 -0.78864900    0.58868900   -0.33239300
 C                 -1.06884800   -0.89429900   -0.29434600
 C                  0.54804600    1.07101100   -0.26685400
 O                  0.06729000   -1.68957300   -0.03367500
 O                 -2.06699300   -1.11018800    0.67742100
 H                 -1.42948100   -1.22551300   -1.27435100
 O                 -1.66089700    1.30486100   -1.17655900
 C                  1.64041300    0.08403900    0.09568700
 H                  0.84340500    1.91834500   -0.87862300
 C                  1.01253200   -1.08492100    0.84163100
 H                 -2.34488100   -2.03028100    0.62166300
 H                 -2.50508900    1.38927300   -0.72671500
 O                  2.25208500   -0.35502100   -1.09864400
 H                  2.40302300    0.57316000    0.69916300
 H                  1.76298400   -1.84516900    1.05445000
 H                  0.53931300   -0.75708800    1.77203800
 H                  1.67284700   -1.02189000   -1.48735400
 H                 -0.63938100    1.26367400    0.94333000
 O                  0.33082200    1.88707300    1.34432300
 H                  0.25784200    2.85115500    1.31111400

\end{verbatim}
\paragraph{\ce{TS-AXP_{3-4}}}
\begin{verbatim}
0 1
 C                 -1.28323900   -0.24229000   -0.07611200
 C                 -0.56929800   -1.10097800   -1.12364100
 C                 -0.52614100    0.35399100    0.94970500
 O                  0.81166700   -1.28080100   -0.85122000
 H                 -1.01839700   -2.09457300   -1.09773300
 H                 -0.68938400   -0.68232400   -2.12700700
 O                 -2.58026900   -0.64789000    0.26655600
 C                  0.98136100    0.24292700    0.94881200
 H                 -1.01586900    0.62486800    1.87795500
 C                  1.48546600   -0.08931500   -0.45948700
 H                 -3.19555900   -0.35460500   -0.40982400
 O                  1.28510600   -0.81664600    1.83090200
 H                  1.44191100    1.14898300    1.33631600
 O                  1.33376400    0.93969500   -1.36653600
 H                  2.53483000   -0.38422200   -0.40930700
 H                  1.14418100   -1.63984100    1.34480100
 H                  0.70640500    1.60410100   -1.01912800
 H                 -1.28063300    1.01001100   -0.40938400
 O                 -0.63583000    2.14791100    0.01856700
 H                 -0.87187900    3.04342500    0.29148500

\end{verbatim}
\paragraph{\ce{TS-AXP_{4-3}}}
\begin{verbatim}
0 1
 C                  0.94218400   -0.56312100   -0.62225800
 C                 -0.33092800   -1.40187700   -0.44015400
 C                  0.70713600    0.84647300   -0.12556400
 O                 -0.64320300   -1.36595300    0.95030200
 O                 -1.38690000   -1.03071300   -1.24994500
 H                 -0.10802700   -2.44771600   -0.64835500
 O                  2.02155100   -1.15665400    0.08767800
 H                  1.23749300   -0.55257000   -1.67243300
 C                 -0.29600400    1.04099600    0.85944000
 O                  1.83042000    1.67463600   -0.11757100
 C                 -1.12494600   -0.10929200    1.37950800
 H                 -1.57837200   -0.07740400   -1.18237000
 H                  1.67287200   -1.45270400    0.93778100
 H                 -0.26952900    1.95167800    1.44713400
 H                  2.56979100    1.13350700    0.18585100
 H                 -2.17761400    0.01678600    1.12283100
 H                 -1.03380000   -0.10527300    2.46956400
 H                 -0.21763000    1.43131000   -0.85125100
 O                 -1.49354700    1.72413900   -0.59359900
 H                 -2.10639000    2.45967500   -0.72950600

\end{verbatim}
\paragraph{\ce{TS-AXP_{4-5}}}
\begin{verbatim}
0 1
 C                  0.40255100    0.27980300    1.13438700
 C                 -0.17784900    1.24292200    0.09149000
 C                  1.14389000   -0.86234300    0.42668100
 C                 -0.83465700    0.45738400   -1.02295900
 O                  0.79459600    2.06944600   -0.51225900
 H                 -0.90499100    1.89447400    0.57275000
 O                 -0.58260700   -0.23603200    1.98479100
 H                  1.12831700    0.82780000    1.73997400
 O                  2.20614700   -0.27124900   -0.29831400
 O                  0.31730400   -1.60994400   -0.42165800
 H                  1.51477700   -1.58097900    1.15910200
 C                 -0.51287700   -0.88438000   -1.29585600
 H                 -1.31616500    1.03006900   -1.80617000
 H                  1.56219900    1.51876300   -0.71730600
 H                 -1.39482600   -0.38036000    1.46755400
 H                  2.68673100   -0.96402000   -0.76146700
 H                 -0.45846500   -1.22952200   -2.32252300
 H                 -1.74869200   -0.99414300   -0.85423900
 O                 -2.46266100   -0.16095600   -0.06186300
 H                 -3.37746400    0.14748700   -0.00570500
\end{verbatim}
\subsubsection{Ring contractions}
\paragraph{TC-C-OH2-c}
\begin{verbatim}
0 1
 C                  0.07653500    0.84888000   -0.20689300
 C                  1.51995000    0.47578100    0.13927800
 C                 -0.50161500   -0.54454200   -0.49716800
 C                  1.32245600   -0.81377000    0.92016500
 O                  2.25302500    0.16422800   -1.03059600
 H                  1.99851000    1.25776200    0.73441200
 O                 -0.47958500    1.44990000    0.93777100
 H                  0.01550300    1.50494100   -1.08124600
 C                 -1.98446400   -0.66025600   -0.23084400
 O                  0.19033900   -1.46089000    0.33277900
 H                 -0.32617100   -0.79731700   -1.55320100
 H                  2.18906100   -1.46864600    0.82991700
 H                  1.11282200   -0.59768000    1.97040000
 H                  2.40916800    0.97129400   -1.52789600
 H                 -1.41202500    1.62125600    0.76584200
 O                 -2.74550300    0.26294800   -0.36250800
 H                 -2.33024400   -1.65765300    0.09496800
\end{verbatim}
\paragraph{TC-C-OH2-t}
\begin{verbatim}
0 1
 C                  0.43024700   -0.71961100    0.56407400
 C                 -0.38000300   -0.65130200   -0.72998500
 C                  1.77039200   -0.00408000    0.43859100
 C                 -0.97182800    0.75894300   -0.58937900
 O                 -1.38863700   -1.63320500   -0.78288300
 H                  0.24327900   -0.76976600   -1.61678700
 O                 -0.39830200   -0.10760600    1.53989900
 H                  0.62201100   -1.75482100    0.85972600
 O                  2.59514000   -0.32526600   -0.37086800
 H                  1.93089200    0.81623200    1.16287300
 C                 -1.29967600    0.80844100    0.89870200
 O                  0.02537100    1.74057700   -0.81521800
 H                 -1.84697800    0.90155800   -1.22784600
 H                 -1.77857600   -1.73065300    0.09187100
 H                 -1.14545600    1.81435300    1.29097700
 H                 -2.32505100    0.48998900    1.10347000
 H                  0.33651100    1.68276800   -1.72373300

\end{verbatim}
\paragraph{TS-TC-C-OH2-c}
\begin{verbatim}
0 1
 C                  1.60322100   -0.32679000    0.30022400
 C                  1.01533400   -1.69104600   -0.06039900
 C                 -0.98924500   -0.66332200   -0.25394600
 C                 -0.85900300    0.64892000    0.32974400
 C                  0.69765600    0.76131800   -0.25366600
 H                  1.09654900   -1.86307400   -1.13896000
 H                  1.47609100   -2.51019000    0.48342300
 H                  1.65023600   -0.21995200    1.39078200
 H                 -1.19314400   -0.75883300   -1.32239400
 H                 -0.68552700    0.56910600    1.42227300
 H                  0.67606400    0.71809200   -1.35098500
 O                 -0.38635800   -1.68432500    0.33802700
 O                 -3.13752500   -0.57004200   -0.15419800
 O                 -1.61464600    1.61530300   -0.11832400
 H                 -2.95868800    0.40791700   -0.20038700
 O                  1.12827500    2.01426300    0.17410600
 H                  0.37722100    2.61387800    0.02775100
 O                  2.89147800   -0.27725800   -0.26945100
 H                  3.22652500    0.61866900   -0.15620300
 H                 -3.52290600   -0.73361500    0.71167700
\end{verbatim}
\paragraph{TS-TC-C-OH2-t}
\begin{verbatim}
0 1
 C                  1.45883800   -0.81299500    0.35515500
 C                  0.74184700    0.30812000   -0.30030700
 C                 -0.58762900    0.78035000    0.18631700
 C                 -1.61272600   -0.28135100   -0.22669300
 C                 -0.97953100   -1.64425500    0.05296200
 O                  2.66129400   -0.96117900    0.06365600
 O                  1.88230000    1.64522100    0.27882000
 O                 -0.85598600    2.04087600   -0.38624600
 O                 -2.79606500    0.01767200    0.48778700
 O                  0.27518700   -1.63884800   -0.56815900
 H                  1.08541300   -1.08556500    1.35888500
 H                  0.97859100    0.43871100   -1.34975500
 H                 -0.57958200    0.84353600    1.28209000
 H                 -1.77060100   -0.17854900   -1.30765300
 H                 -0.91150400   -1.79935800    1.14100200
 H                  2.68562700    1.08811500    0.17813200
 H                  1.92127500    2.39603700   -0.32527400
 H                 -1.75050400    2.28947200   -0.12990300
 H                 -3.53255500   -0.46817900    0.10880800
 H                 -1.58479100   -2.45337200   -0.36780300

\end{verbatim}

\subsection{Xylose decompositions}
\subsubsection{Dehydration}
\paragraph{A1}
\begin{verbatim}
0 1
 C                  1.07374400   -0.23077000   -0.38595400
 C                  1.74260800    0.52215700    0.76711700
 C                 -1.76563400    0.93129300   -0.40731800
 C                 -1.42411800   -0.42702100    0.09693900
 C                 -0.17798100   -0.91750900    0.07879000
 H                  1.10271800    1.34320500    1.10691500
 H                  1.90154400   -0.16622800    1.59966700
 H                  0.86775200    0.48090600   -1.19626400
 H                 -0.93045400    1.48629300   -0.86888500
 H                  0.00456500   -1.91602700    0.47200600
 O                  3.02728500    0.97807700    0.37508600
 O                 -2.86152600    1.41182900   -0.32740700
 H                  2.93829500    1.79292300   -0.12584000
 O                 -2.50583800   -1.08454800    0.57176000
 H                 -2.24917600   -1.96195600    0.87083100
 O                  1.94501900   -1.23851600   -0.85676400
 H                  2.83351700   -0.86275000   -0.85728500

\end{verbatim}
\paragraph{A2}
\begin{verbatim}
0 1
 C                 -1.39063900    0.08367200    0.71633200
 C                 -1.09929000    1.43157500    0.06403800
 C                  1.95332300    0.10304100   -0.25257700
 C                  0.79177800   -0.86619500   -0.00789000
 C                 -0.09948700   -0.61495500    1.17993800
 H                 -2.04294600    1.94925100   -0.13413700
 H                 -0.48534700    2.05271000    0.71936200
 H                 -2.02821700    0.24043900    1.59189500
 H                  2.42921500   -0.00561400   -1.24253900
 H                 -0.35660300   -1.57848300    1.62507800
 O                 -0.36528900    1.25151800   -1.13449500
 O                  2.31662400    0.88376400    0.57887900
 H                 -0.92716800    0.75260400   -1.73847400
 O                  0.61183700   -1.77400400   -0.78432200
 H                  0.41160500    0.00772300    1.91478600
 O                 -2.10927700   -0.71005100   -0.21374700
 H                 -1.56580700   -1.45127900   -0.50553000

\end{verbatim}
\paragraph{A2-a1}
\begin{verbatim}
0 1
 C                  1.00249400   -0.59809700    0.80577900
 C                  2.00128200   -0.04694900   -0.20531000
 C                 -0.93108000    0.70034200    0.63293800
 C                 -1.18388800   -0.61237200   -0.14137600
 C                 -0.02130300   -1.53685500    0.16839100
 H                  2.53650700    0.79360300    0.25145900
 H                  2.72784000   -0.83050300   -0.43480700
 H                  1.54388900   -1.05259400    1.63598300
 H                 -1.75874400    0.93724300    1.30518000
 H                 -0.36616500   -2.29023700    0.88136500
 O                  1.40385000    0.31245000   -1.42934800
 O                 -0.75202600    1.72997300   -0.31039500
 H                  0.81591000    1.06579000   -1.28626000
 O                 -2.14058100   -0.82509100   -0.82214700
 H                  0.32033100   -2.02577600   -0.74225900
 O                  0.20738100    0.45630400    1.39715200
 H                 -0.77358500    2.57698000    0.14470200

\end{verbatim}
\paragraph{A2-a2}
\begin{verbatim}
0 1
 C                 -0.92153400    0.94382500    0.19736200
 C                 -1.52800300   -0.05724300   -0.79430300
 C                  0.09391200   -0.88082400    0.55766600
 C                  1.20626800   -0.00328500   -0.04310000
 C                  0.48622800    1.33193600   -0.27625100
 H                 -2.57555600   -0.26920900   -0.57578500
 H                 -1.41071200    0.24431800   -1.83631000
 H                 -1.57103600    1.74977200    0.52406500
 H                  0.38372100   -1.75243100    1.13425600
 H                  0.92442700    2.09809900    0.36561800
 O                 -0.74321800   -1.23089700   -0.53512100
 O                  2.34015800   -0.29274200   -0.27946700
 H                  0.55971000    1.64557900   -1.31748600
 O                 -0.63841300    0.05831600    1.29676300
\end{verbatim}
\paragraph{A2-b1}
\begin{verbatim}
0 1
 C                  0.00000000    0.50947200    0.00000000
 C                 -0.71407300    1.63550800    0.00000000
 C                 -0.67145700   -0.81193000    0.00000000
 H                 -1.79415100    1.57738800    0.00000000
 H                 -0.24084700    2.60801400    0.00000000
 H                 -1.77357800   -0.80643900    0.00000000
 O                 -0.02963400   -1.83275900    0.00000000
 O                  1.34425800    0.46866100    0.00000000
 H                  1.60476500   -0.46447600    0.00000000
\end{verbatim}
\paragraph{HAA}
\begin{verbatim}
0 1
 C                  0.66684700    0.64812100    0.00006400
 H                  0.93666900    1.24427600    0.88256300
 C                 -0.83057400    0.48311700    0.00001300
 H                 -1.43161800    1.41039000    0.00008800
 O                  1.33375100   -0.57601400    0.00003700
 H                  0.93669200    1.24435300   -0.88238100
 O                 -1.34870200   -0.60116900   -0.00010800
 H                  0.66022900   -1.26898500   -0.00016000
\end{verbatim}
\paragraph{MGO}
\begin{verbatim}
0 1
 C                  0.00000000    0.55500000    0.00000000
 O                 -0.58488900    1.60619900    0.00000000
 C                  1.48981000    0.38817400    0.00000000
 H                  1.96932200    1.36407800    0.00000000
 H                  1.79273600   -0.19073800    0.87548200
 H                  1.79273600   -0.19073800   -0.87548200
 C                 -0.84342900   -0.72906700    0.00000000
 H                 -1.93468200   -0.55101900    0.00000000
 O                 -0.35241000   -1.82072800    0.00000000

\end{verbatim}
\paragraph{A2-c1}
\begin{verbatim}
0 1
 C                 -1.74821400    0.11154200    0.49433600
 C                 -1.35272000    1.20800600   -0.50235300
 C                  1.98173400    0.39374600    0.40046900
 C                  0.55731100   -0.01384100    0.03634500
 C                 -0.41278100   -0.14198700    1.20707200
 H                 -1.88075400    1.08630200   -1.44856000
 H                 -1.53219600    2.20970600   -0.10461600
 H                 -2.52718000    0.45067200    1.17633500
 H                  2.06846800    1.29793300    1.02844000
 H                 -0.21574300    0.64946100    1.93535500
 O                  0.06205200    1.06573200   -0.72076600
 O                  2.93592100   -0.20516400   -0.00518300
 H                  1.40009300   -1.29144500   -1.14709600
 O                  0.55696900   -1.21001900   -0.68139200
 H                 -0.36669200   -1.11746300    1.68700200
 O                 -2.25018600   -1.04743800   -0.13079300
 H                 -1.53601700   -1.45484700   -0.63700200

\end{verbatim}
\paragraph{A2-c2}
\begin{verbatim}
0 1
 C                 -2.07089000    0.40597000   -0.30239400
 C                 -1.80532100   -0.84687900    0.03818900
 C                  1.43159600   -0.17233400   -0.76899900
 C                  0.23792900    0.12730100    0.13779700
 C                 -0.76506100    1.14942300   -0.43181300
 H                 -2.46458600   -1.68408000    0.21233200
 H                 -3.05106700    0.82511100   -0.46273400
 H                  1.18271200   -0.42004600   -1.81618800
 H                 -0.54278200    1.40137700   -1.47401800
 O                 -0.47055700   -1.12548500    0.18799100
 O                  2.55501600   -0.17487100   -0.35524600
 H                  1.52290400    0.12684000    1.55538200
 O                  0.65233800    0.51376800    1.39063800
 H                 -0.71107500    2.06261500    0.16147800

\end{verbatim}
\paragraph{A2-c3}
\begin{verbatim}
0 1
 C                 -1.33911400    1.14126200    0.23268800
 C                 -1.51662800   -0.37482400    0.43563300
 C                 -0.07490100   -0.81535900    0.49971000
 C                  0.67682300    0.18938000    0.04350000
 C                  2.14076700    0.22488900   -0.20667400
 H                 -2.08656100   -0.58679100    1.33970900
 H                 -2.03703300    1.54898900   -0.49679500
 H                  2.52142500    1.18209300   -0.60496600
 H                  0.28279400   -1.79917300    0.76084900
 O                  0.00361200    1.33454100   -0.24946100
 O                  2.85899500   -0.71316100    0.00662500
 O                 -2.22717900   -1.01629200   -0.61005700
 H                 -1.65967700   -1.05383800   -1.38639600
 H                 -1.42605000    1.67592600    1.18160500

\end{verbatim}
\paragraph{FF}
\begin{verbatim}
0 1
 C                  0.00000000    0.28292200    0.00000000
 C                 -0.95583200   -0.68892200    0.00000000
 C                 -0.06686500    1.74305500    0.00000000
 C                 -0.25635000   -1.92961800    0.00000000
 H                 -2.01994400   -0.51674900    0.00000000
 O                  1.23922900   -0.27898200    0.00000000
 O                 -1.10390200    2.35642600    0.00000000
 H                  0.91526200    2.25023000    0.00000000
 C                  1.06737700   -1.61144900    0.00000000
 H                 -0.67366700   -2.92295100    0.00000000
 H                  1.96575500   -2.20600900    0.00000000
\end{verbatim}
\paragraph{TS-A}
\begin{verbatim}
0 1
 C                 -1.17826900    0.20567100   -0.26499100
 C                 -1.50313500   -1.07585400    0.50378100
 C                  1.66491400   -0.87563300   -0.73284500
 C                  1.39660300    0.22971100    0.13283700
 C                  0.07021300    0.84584700    0.28430400
 H                 -0.74326000   -1.84157900    0.32966700
 H                 -1.54299900   -0.85724000    1.57331900
 H                 -1.04109400   -0.02764600   -1.32962700
 H                  0.97286900   -1.02301700   -1.58366200
 H                  1.32641800    1.60948700   -0.70063000
 H                 -0.09455400    1.22996300    1.29107100
 O                 -2.80002200   -1.52107100    0.13687200
 O                  2.64517200   -1.59708200   -0.59189800
 H                 -2.73767800   -2.07701600   -0.64377700
 O                  2.20024500    0.32543400    1.27100000
 H                  2.86980200   -0.36617900    1.16722100
 O                  0.31435600    2.12222000   -0.62513300
 H                  0.24674900    2.97707500   -0.16847600
 O                 -2.22308000    1.13925000   -0.10461500
 H                 -3.05158300    0.64769500   -0.16342900
\end{verbatim}
\paragraph{TS-A1A2}
\begin{verbatim}
0 1
 C                 -1.08810100    0.31477100   -0.29293400
 C                 -1.75421300   -0.73846000    0.59079800
 C                  1.93022500   -0.97472500   -0.34326600
 C                  1.49633900    0.41547600    0.05075500
 C                  0.17695900    0.85355500    0.32001300
 H                 -1.10019200   -1.60935200    0.71417600
 H                 -1.94613100   -0.30478000    1.57474000
 H                 -0.86085900   -0.14087300   -1.27082700
 H                  1.15739200   -1.57173300   -0.86405300
 H                  0.07583600    1.20798900    1.34979300
 O                 -3.02081400   -1.10458500    0.06569500
 O                  3.00807400   -1.41725000   -0.07525600
 H                 -2.90047100   -1.73218200   -0.65149300
 O                  2.32557800    1.36505100    0.17737000
 H                  1.14102700    1.94338700    0.20865600
 O                 -1.97272300    1.40127900   -0.47098700
 H                 -2.85477500    1.02787600   -0.58776500
\end{verbatim}
\paragraph{TS-A2a1}
\begin{verbatim}
0 1
 C                 -0.76752100   -0.51230300   -0.29628600
 C                 -2.07192700   -0.45272700    0.47607600
 C                  1.29866600    0.94834500   -0.31228300
 C                  1.61795000   -0.52907200   -0.06374000
 C                  0.36670900   -1.20494700    0.46560900
 H                 -2.35179600   -1.47874500    0.74843900
 H                 -1.90912300    0.10643800    1.40142300
 H                 -0.92205900   -0.99021800   -1.26791700
 H                  1.73601600    1.37454200   -1.22218300
 H                  0.40467100   -2.28177000    0.31481000
 O                 -3.10144600    0.21105500   -0.21253200
 O                  1.16246500    1.68751700    0.75984900
 H                 -3.33502000   -0.28619100   -1.00059800
 O                  2.68460200   -1.03209700   -0.26491400
 H                  0.29736300   -0.98871100    1.53724900
 O                 -0.30414700    0.83101500   -0.59065600
 H                 -0.11510800    1.46896400    0.33855300

\end{verbatim}
\paragraph{TS-A2a12}
\begin{verbatim}
0 1
 C                 -1.47280400   -0.42141100    0.40608200
 C                 -0.97765300   -1.38936500   -0.69387000
 C                  0.56678000    0.31178500    1.02503800
 C                  0.11685600    1.22922700   -0.10316900
 C                 -1.38638700    0.98925200   -0.14965100
 H                 -0.80015800   -2.37698300   -0.23872100
 H                 -1.80612700   -1.49370800   -1.41374700
 H                 -2.36988500   -0.70460000    0.94882900
 H                  1.42244400    0.42420800    1.67406000
 H                 -1.88207600    1.71317200    0.50400200
 O                  0.12872700   -0.79755900   -1.21812500
 O                  2.18216100   -1.00999700    0.12226800
 H                  1.43043900   -1.08498400   -0.57408200
 O                  0.79198000    2.07561400   -0.61011700
 H                 -1.75558400    1.07637300   -1.16772500
 O                 -0.38421500   -0.42591000    1.41687700
 H                  2.93097500   -0.60758900   -0.32642200
\end{verbatim}
\paragraph{TS-A2b1}
\begin{verbatim}
0 1
 C                 -1.29032900   -0.58478400    0.32534800
 C                 -1.42065700    0.92810200    0.23718600
 C                  1.96903000    0.80325200    0.35442000
 C                  1.21373000   -0.50679900    0.23363300
 C                  0.44781200   -0.94637000    1.30367900
 H                 -1.29219400    1.39402600    1.21530700
 H                 -0.67089100    1.32041900   -0.46122400
 H                 -1.86749300   -1.04763200    1.13444100
 H                  1.94455700    1.26390400    1.36104100
 H                  0.20259200   -1.99852400    1.36561600
 O                 -2.72910500    1.23068800   -0.19162900
 O                  2.56130000    1.29541200   -0.55866500
 H                 -2.85727100    0.83007400   -1.05786700
 O                  1.16484000   -1.06906800   -0.91310200
 H                  0.49168800   -0.39946700    2.23868000
 O                 -1.18917100   -1.23518800   -0.77015300
 H                  0.06859000   -1.29796500   -1.05319300
\end{verbatim}
\paragraph{TS-MGO}
\begin{verbatim}
0 1
 C                  0.40194300    0.00908800   -0.05509700
 C                  1.59557700   -0.73791500    0.08589000
 C                 -1.00438800   -0.52930100   -0.16431800
 H                  1.70766500   -1.74351000   -0.30246900
 H                  2.07621400   -0.56512100    1.04883000
 H                 -1.07472100   -1.55711400   -0.56446100
 O                 -1.96057800    0.09039400    0.19761200
 O                  0.64954100    1.25066200   -0.09637100
 H                  1.82034400    0.68606600   -0.19067300

\end{verbatim}
\paragraph{TS-A2c1}
\begin{verbatim}
0 1
 C                 -1.35353100    0.45375900   -0.14862900
 C                 -1.22218400   -0.92010200   -0.80534200
 C                  1.90154000    0.59504400   -0.32468100
 C                  0.89576700   -0.29312200    0.41115000
 C                 -0.31935300    0.41366800    0.98338600
 H                 -1.70445300   -1.67162000   -0.17478600
 H                 -1.63064300   -0.95625900   -1.81371600
 H                 -1.07426500    1.22830500   -0.87284700
 H                  2.78177500    0.04513300   -0.70370600
 H                 -0.05107800    1.41336800    1.32702400
 O                  0.19231100   -1.13974600   -0.85594500
 O                  1.75225100    1.77549400   -0.46307100
 H                  0.68370000   -1.84568700   -0.07148300
 O                  1.38189600   -1.37597400    0.95639800
 H                 -0.70655600   -0.18161800    1.81196800
 O                 -2.68453400    0.60825200    0.27601800
 H                 -2.84730600    1.52869100    0.49505000
\end{verbatim}
\paragraph{TS-A2c12}
\verbatiminput{SI/geometries/tsa2c12.xyz}
\paragraph{TS-A2c13}
\verbatiminput{SI/geometries/tsa2c13.xyz}
\paragraph{TS-A2c2FF}
\verbatiminput{SI/geometries/tsa2c2ff.xyz}
\paragraph{TS-A2c3FF}
\verbatiminput{SI/geometries/tsa2c3ff.xyz}
\subsubsection{Cyclization}
\paragraph{B1}
\verbatiminput{SI/geometries/b1.xyz}
\paragraph{ADX}
\verbatiminput{SI/geometries/adx.xyz}
\paragraph{TS-B}
\verbatiminput{SI/geometries/tsb.xyz}
\paragraph{TS-B1ADX}
\verbatiminput{SI/geometries/tsb1adx.xyz}
\subsubsection{\ce{C-C} scission}
\paragraph{C1}
\verbatiminput{SI/geometries/c1.xyz}
\paragraph{C2}
\verbatiminput{SI/geometries/c2.xyz}
\paragraph{FD}
\begin{verbatim}
0 1
 O                  0.00000000    0.00000000    0.67105800
 C                  0.00000000    0.00000000   -0.52560500
 H                  0.00000000    0.93942400   -1.10741700
 H                  0.00000000   -0.93942400   -1.10741700

\end{verbatim}
\paragraph{TS-C}
\verbatiminput{SI/geometries/tsc.xyz}
\paragraph{TS-C2C1}
\verbatiminput{SI/geometries/tsc2c1.xyz}
\paragraph{TS-C1HAA}
\verbatiminput{SI/geometries/tsc1haa.xyz}
\subsubsection{Isomerisation}
\paragraph{D1}
\verbatiminput{SI/geometries/d1.xyz}
\paragraph{D1-a1}
\verbatiminput{SI/geometries/d1a1.xyz}
\paragraph{D1-a2}
\verbatiminput{SI/geometries/d1a2.xyz}
\paragraph{D1-a3}
\verbatiminput{SI/geometries/d1a3.xyz}
\paragraph{D1-b1}
\verbatiminput{SI/geometries/d1b1.xyz}
\paragraph{AD}
\verbatiminput{SI/geometries/ad.xyz}
\paragraph{ETH}
\verbatiminput{SI/geometries/eth.xyz}
\paragraph{D1-c1}
\verbatiminput{SI/geometries/d1c1.xyz}
\paragraph{DHA}
\verbatiminput{SI/geometries/dha.xyz}
\paragraph{D1-d1}
\verbatiminput{SI/geometries/d1d1.xyz}
\paragraph{D1-d2}
\verbatiminput{SI/geometries/d1d2.xyz}
\paragraph{D1-d3}
\verbatiminput{SI/geometries/d1d3.xyz}
\paragraph{D1-d4}
\verbatiminput{SI/geometries/d1d4.xyz}
\paragraph{D1-d5}
\verbatiminput{SI/geometries/d1d5.xyz}
\paragraph{D1-d6}
\verbatiminput{SI/geometries/d1d6.xyz}
\paragraph{D1-d7}
\verbatiminput{SI/geometries/d1d7.xyz}
\paragraph{DFO}
\verbatiminput{SI/geometries/dfo.xyz}